\documentclass[a4paper,fleqn,usenatbib,useAMS]{mnras}

\usepackage{graphicx}	% Including figure files
\usepackage{amsmath}	% Advanced maths commands
\usepackage{amssymb}	% Extra maths symbols
\usepackage{multicol}        % Multi-column entries in tables
\usepackage{bm}		% Bold maths symbols, including upright Greek
\usepackage{pdflscape}	% Landscape pages
\usepackage{rotating}
\usepackage{tabularx}
\usepackage{ragged2e}
\usepackage{lipsum}% for dummy text
\usepackage{booktabs} %for table bold lines
\usepackage[normalem]{ulem} % to be removed at the end of correction!!!!!!!!!!!
\newcommand{\nbody}{\texttt{NBODY6}}
\newcommand{\nbodyplus}{\texttt{NBODY6++}}
\newcommand{\nb}{\texttt{NBODY6++GPU}}
\newcommand{\nbseven}{\texttt{NBODY7}}

\newcommand{\mpi}{\texttt{MPI}}

\newcommand{\bh}{IMBH}
\newcommand{\sbh}{SMBH}
\newcommand{\ffp}{FFP}
\newcommand{\bfp}{BFP}
\newcommand{\bbp}{BP}
\newcommand{\nhs}{NHS}

\newcommand{\nstars}{N_{\rm s}}

\newcommand{\kms}{\,km\,s$^{-1}$}

\newcommand{\msun}{M_\odot}
\newcommand{\maverage}{\langle m \rangle}
\newcommand{\mbh}{M_{\rm BH}}
\newcommand{\mstar}{M_{*}}

\newcommand{\mcluster}{M_{\rm cl}}
\newcommand{\rcluster}{R_{\rm cl}}

\newcommand{\rvir}{r_{\rm vir}}
\newcommand{\rtidal}{r_{\rm tidal}}
\newcommand{\halfmass}{r_{\rm hm}}

\newcommand{\crossingtime}{t_{\rm cr}}
\newcommand{\relaxationtime}{t_{\rm rh}}

\usepackage[T1]{fontenc}
\usepackage{ae,aecompl}

\title[Planets in star clusters II: IMBH and planets]{Planetary Systems in a Star Cluster II: intermediate-mass black holes and planetary systems}

\author[Flammini Dotti, Kouwenhoven, Shu, Hao \& Spurzem]{Francesco Flammini Dotti$^{1,2}$,
M.B.N. Kouwenhoven$^{1}$\thanks{Contact e-mail: \href{mailto:t.kouwenhoven@xjtlu.edu.cn}{t.kouwenhoven@xjtlu.edu.cn}},
Qi Shu$^{3,4}$, Wei Hao$^{5}$
\newauthor{and Rainer Spurzem$^{6,7,3}$\thanks{Research  Fellow  at  Frankfurt  Institute  for  Advanced  Studies (FIAS)}}
\\
$^{1}$Department of Physics, School of Science, Xi{'}an Jiaotong-Liverpool University, 111 Ren{'}ai Rd., \\
Suzhou Dushu Lake Science and Education Innovation District, Suzhou Industrial Park, Suzhou 215123, P.R. China\\
$^{2}$Department of Mathematical Sciences, University of Liverpool, Liverpool L69 3BX, UK\\
$^3$Kavli Institute for Astronomy and Astrophysics at Peking University, 5 Yiheyuan Rd., Haidian District, 100871, Beijing, China\\
$^4$Department of Astronomy, School of Physics, Peking University, Yiheyuan Lu 5, Haidian Qu, 100871, Beijing, China\\
$^5$Max Planck Institut f\"{u}r Astrophysik, Karl-Schwarzschild-Str. 1 85741 Garching, Germany \\
$^6$National Astronomical Observatories and Key Laboratory of Computational Astrophysics, \\
Chinese Academy of Sciences, 20A Datun Rd., Chaoyang District, 100012, Beijing, China\\
$^7$Astronomisches Rechen-Institut, Zentrum f\"ur Astronomie, University of Heidelberg, M\"onchhofstrasse 12--14, 69120, Heidelberg, Germany}

\date{Last updated \today}
\pubyear{2020}

\begin{document}
\label{firstpage}
\pagerange{\pageref{firstpage}--\pageref{lastpage}}
\maketitle

\bibliographystyle{mnras}

\begin{abstract}
Most stars form in dense stellar environments. It is speculated that some dense star clusters may host intermediate-mass black holes (\bh{}s), which may have formed from runaway collisions between high-mass stars, or from the mergers of less massive black holes. Here, we numerically explore the evolution of populations of planets in star clusters with an \bh{}. We study the dynamical evolution of single-planet systems and free-floating planets, over a period of 100~Myr, in star clusters without an \bh{}, and in clusters with a central \bh{} of mass $100~\msun$ or $200~\msun$.  In the central region ($r\la 0.2$~pc), the \bh{}'s tidal influence on planetary systems is typically 10~times stronger than the average neighbour star. For a star cluster with a $200\msun$ \bh{}, the region in which the \bh{}'s influence is stronger within the virial radius ($\sim 1$~pc). 
The \bh{} quenches mass segregation, and the stars in the core tend to move towards intermediate regions. The ejection rate of both stars and planets is higher when an \bh{} is present. The rate at which planets are expelled from their host star rate is higher for clusters with higher \bh{} masses, for $t<0.5\relaxationtime$, while remains mostly constant while the star cluster fills its Roche lobe, similar to a star cluster without an \bh. The disruption rate of planetary systems is higher in initially denser clusters, and for wider planetary orbits, but this rate is substantially enhanced by the presence of a central \bh{}.
\end{abstract}

% Select between one and six entries from the list of approved keywords. Don't make up new ones.


%%%%%%%%%%%%%%%%%%%%%%%%%%%%%%%%%%%%%%%%%%%%%%%%%%
%%%%%%%%%%%%%%%%% BODY OF PAPER %%%%%%%%%%%%%%%%%%

\section{Introduction} \label{sec:intro}

Planetary systems are believed to be common in star clusters \citep[e.g.,][]{mayo2018, thompson2018}. In order to understand the origin and dynamical evolution of planetary systems and rogue planets in star clusters, it is necessary to carefully study the effect of their environments, i.e., that of the star-forming region in which they were born \citep[e.g.,][]{lada2003} and in which they remain the first million years, and that of the Galactic tidal field, the open cluster, or the globular cluster in which they may spend the billions of years that follow.

In this study we analyse the effect of a central \bh{} on the evolution of planetary systems and on populations of free-floating planets in star clusters. The differential gravitational force of a central \bh{} affects the evolution of both binary star systems and planetary systems. A massive \bh, located in or near the centre of the star cluster affects not only the densest regions of the star cluster, but when sufficiently massive, it can dominate the evolution of planetary systems throughout the entire star cluster. In order to understand the evolution of planetary systems in star clusters, it is thus interesting and necessary to quantify the effect of a central \bh{} on a planetary system, and to compare this to the contribution of neighbouring stars.

Many of the currently known black holes have been identified through their gravitational influence on neighbouring objects or through accretion of gas. Unlike stellar-mass black holes \citep[e.g., A0620-00/V616,][]{cantrell2010} and supermassive black holes \citep[e.g., SgrA*,][]{schodel2002}, \bh{}s have not yet been directly observed. Several candidate \bh{}s have been identified in star clusters. These include HLX-1 in ESO\,243-49 \citep{farrell2009}, and the candidate \bh{}s in NGC\,4395 \citep[][although its mass is between that of a \sbh{} and an \bh]{denbrok2015}, 47\,Tuc \citep{koliopanos2017}, and NGC\,6624 \citep{perera2017}. \cite{koliopanos2017} estimate that globular clusters can host \bh{} up to $\sim 1000~\msun$. A promising \bh{} candidate was recently identified by \cite{lin2020}. The \bh{} candidate was identified through a stellar tidal disruption event, and is estimated to have a mass of $50\,000\msun$.  Gravitational waves signatures of \bh{} candidates in star clusters may substantially increase the number of known \bh{} candidates in star clusters in the near future \citep[e.g.,][]{ligo2017}.

The existence of \bh{}s in star clusters is predicted by several theoretical models. Among these, the model involving a stellar runaway collision is the least controversial. When the local stellar density in the centre of a star cluster is sufficiently high, runaway stellar collisions can ultimately result in the formation of an \bh{} within roughly five relaxation times \citep[e.g.,][]{pz2002, pz2004a, gurkan2004, freitag2006}. The \bh{}s formed through this channel typically have masses of $\sim 10^2\msun$.
Mergers of stellar-mass black holes can also result in \bh{}s, with masses of up to $\sim 10^5~\msun$.  \cite{miller2002} show that more massive black holes ($>50\msun$) sink towards a a star cluster's centre and merge to form an \bh{}, while less massive black holes may escape within $\sim 10^7$ Myr \citep[e.g.,][]{sigurdsson1993, kulkarni1993, devita2018}.
Population~III stars are thought to have had substantially higher masses than Population~I and~II stars. It is possible to obtain black holes with masses over $200~\msun$ \citep[e.g.,][]{madau2001,ricotti2004,wheeler2011} and likely have a central \bh{} \citep{reinoso2018}. This \bh{} forms faster ($\lesssim 3$~Myr) than any other scenario. 
Finally, the most massive \bh{}s may have formed from primordial stars of masses $\sim 10^5~\msun$, in dark matter mini-haloes, at redshift $z > 15$ \citep[e.g.,][]{haiman1996,yoshida2003}.

In a previous study \citep[][]{flammini2020, flammini2019} we explored how the star cluster environment changes the architecture of a complex planetary system (i.e., a multi-planetary system with a range in planetary masses). Strong encounters with neighbouring stars can result in instabilities in a planetary system, which may result in orbital re-configurations or planetary ejections \citep[e.g.,][]{spurzem2009, boley2012}. Even mild perturbations and tidal perturbations can affect short-period planets in the system, resulting in ejection and/or scattering between different planets \citep[e.g.,][]{hao2013, flammini2019}. The effect of a stellar encounter on a planetary system depends on the relative speed of the encountering star, the local stellar density, the effective encounter strength, and the orbital architecture and mass spectrum of the planets in the planetary system.

Upon ejection from their host planetary system, free-floating planets (\ffp{}s) typically obtain a velocity-at-infinity in the range $0.1-10$~\kms{}. The velocity-at infinity tends to be somewhat higher for prompt ejection (as a direct result of star-planet encounters) than for delayed ejection (as a result of a planet-planet scattering); see \cite{malmberg2011}. Once escaped, a \ffp{} may be captured by another star, especially in case of prompt ejection \citep{kouwenhoven2010, malmberg2011, moeckel2011, parker2012, perets2012} or remain in the star cluster, within an estimated time scale of many millions of years \citep[e.g.,][] {hurley2002a, parker2012, craig2013}.

\cite{veras2012} show that the \ffp{} population in the Galactic field is not the result of decaying planetary systems alone, but that close encounters in star clusters are most likely the primary cause of \ffp{}s. Observations of \ffp{}s are limited, but extrapolations suggested that  \ffp{}s are common in the Galactic disk. Estimates range from two Jupiter-mass \ffp{}s per every main-sequence star \citep{sumi2011}, to   $10^5$ \ffp{}s with masses between $10^{-8}\msun$ and $10^{-2}\msun$ per star \citep{strigari2012}. A large contribution of \ffp{} observations comes from the micro-lensing detections of \ffp{} \citep{mao1991, gould1992, abe2004, beaulieu2006, gaudi2012}. Only several \ffp{}s have been unambiguously detected thus far \citep[e.g., CFBDSIR 2149-0403 and PSO J318-22; see][]{delorme2012,liuffp2013}. A recent, and more comprehensive analysis by \cite{mroz2017} finds that \cite{sumi2011} greatly overestimates the abundance of free-floating Jupiter-mass planets in the Galactic field, and also point out that the findings of \cite{sumi2011} 
do not match predictions from planet formation theories \citep{veras2012,ma2016} and that they are inconsistent with surveys of young clusters \citep{scholz2012, pena2012,muzic2015}. Using a much larger sample of microlensing events, \cite{mroz2017} reveal a significantly lower abundance, of roughly one free-floating Jupiter-mass planet or wide-orbit Jupiter-mass planet for every four main-sequence star in the Galactic field.

Earlier works on \bh{}s in star clusters focus on the relation between the stellar dynamics and a central \bh{}. Here, we aim to relate the presence of an \bh{} to the evolution of planetary systems and free-floating debris. We focus on the lower-mass end of the \bh{} mass distribution. We also investigate how the presence of an \bh{} affects the stellar dynamics, and how this contributes to the (in)stability of planetary systems.
This paper is organised as follows. In \S\ref{section:method} we introduce our numerical method and the initial conditions. 
In \S\ref{section:results} we describe our results: we first obtain analytical estimates for the tidal influence of an \bh{} planetary systems; we subsequently analyse disruption rates and escape rates, and how these depend on the properties of the star cluster, the planetary systems, and the central \bh{}. Finally, we summarise and discuss our findings in \S\ref{section:conclusions}.

%%%%%%%%%%%%%%%%%%%%%%%%%%%%%%%%%%%%%%%%%%%%%%%%%%%%%%%%
%%%%%%%%%%%%%%%%%%%%%%%%%%%%%%%%%%%%%%%%%%%%%%%%%%%%%%%%
%%%%%%%%%%%%%%%%%%%%%%%%%%%%%%%%%%%%%%%%%%%%%%%%%%%%%%%%

\section{Methodology and initial conditions} \label{section:method}

\subsection{Initial conditions - star cluster and \bh}\label{ICSC}

\begin{table*}
\caption{Initial conditions for the star clusters: the model identification (column~1, using the syntax C-$Q0$-\bh{} mass), the initial total star cluster mass (column~2), the \bh{} mass (column~3), the initial crossing time and relaxation time (columns~4 and~5), the initial core radius (column~6), the initial half-mass radius (column~7), and other relevant parameters (column~8).  
\label{tab:table}}
\begin{tabular}{lccccccl}
\hline
Model ID  & $\mcluster$ & $M_{\rm IMBH}$ & $\crossingtime$ & $\relaxationtime$ & $r_c$ & $\halfmass$ & Other parameters \\
  &  $\msun$   & $\msun$ & Myr & Myr & pc & pc &  \\
\hline
C50M000 &$ 5.95\times10^3$ & 0 & 0.18 & 27.15 & 0.39 & 0.78 &  \bbp{}=500, \ffp{}=500  \\
C50M100 &$ 6.05\times10^3$ & 100 & 0.18& 27.44 & 0.24 & 0.79 & $a=1-100$~AU, $e=0$  \\
C50M200 &$6.15\times10^3$ & 200 & 0.18 & 27.74 & 0.15 & 0.80 & Ensemble size = 10  \\
\bottomrule
\end{tabular}
\label{table:initialconditions}
\end{table*}

We study open clusters with $N=10\,000$ stellar members. We draw stellar positions and velocities from the  \cite{plummer1911} model, with an initial virial radius of $\rvir=1$~pc. The modeled star clusters have a typical stellar density of $10^4$ stars per cubic parsec, which is higher than that of most known open clusters in the Milky Way, comparable to that of newborn open clusters, and lower than that of globular clusters. Star clusters with different initial stellar densities are discussed in Section~\ref{sec:density}. The clusters are initialized with a virial ratio $Q=1/2$, where $Q = |T/U|$, $T$ is the cluster's total kinetic energy and $U$ the total gravitational energy. Stellar masses are drawn from the \citep{kroupa2001} initial mass distribution  in the mass range $\mstar=0.08-100~\msun$. We adopt the standard solar neighbour tidal field, corresponding to the Solar orbit in the Milky Way. We do not include primordial binaries, and we ignore the presence of any gas remaining from the star-formation process. 
The initial conditions for the reference models are listed in Table~\ref{table:initialconditions}.

The stellar population is initialised with an age of 30~Myr. We use this approach as a more realistic initial condition for \bh{} formation \citep{koliopanos2017}. During the first $\sim 30$~Myr, massive stars \citep[$\ga 25\msun$;][]{fryer1999} evolve into a black hole, favouring the \bh{} formation scenario discussed above. 
Consequently, the $N$-body simulations are initialized with a 30-Myr old stellar population with an evolved mass function. The most massive object at the start of each simulation, excluding the \bh{}, is typically a stellar-mass black hole with a mass between  $50~\msun$ and $70~\msun$. 
Following others studies of \bh{}'s in star clusters, we adopt a stellar metallicity of $Z=0.001$ \citep[e.g,][]{hurley2007, arcasedda2016}.
All models are evolved for $t=100$~Myr, which corresponds to roughly four half-mass relaxation times. In order to enhance statistical significance, we carry out an ensemble of ten realisations for each star cluster model.

We carry out simulations with and without a central \bh{}, to investigate the effect of the presence of an \bh{} on both the stellar and planetary population (see Table~\ref{table:initialconditions}). 

The models contain \bh{}s with masses of $\mbh=0~\msun$ (i.e., no \bh{}), $\mbh=100~\msun$ and $\mbh=200~\msun$. 
The latter two correspond to the lower-mass end of \bh{} mass spectrum, and represent \bh{}s formed through stellar runaway collisions and/or subsequent merging between black holes. 
\bh{}s are initialised at rest at the centre of the star cluster. This is expected because massive objects (the \bh, or its progenitors) tend to sink towards the core and oscillate around the centre of the star cluster \citep[e.g.,][]{miller2002, wrobel2016, arcasedda2016}. 
Note that in our study we do not make attempts to model or explain the formation history the central \bh{}. Forming a central \bh{} through a runaway collision process requires the cluster to evolve for at least a relaxation time \citep[e.g.,][]{pz2002, pz2004a, gurkan2004, freitag2006}, and during this time, the stellar mass function evolves not only due to the process of stellar evolution, but also due to escape events, and the merger events that are responsible for the origin of the \bh{} itself.

%%%%%%%%%%%%%%%%%%%%%%%%%%%%%%%%%%%%%%%%%%%%%%%%%%%%%%%%
%%%%%%%%%%%%%%%%%%%%%%%%%%%%%%%%%%%%%%%%%%%%%%%%%%%%%%%%
%%%%%%%%%%%%%%%%%%%%%%%%%%%%%%%%%%%%%%%%%%%%%%%%%%%%%%%%

\subsection{Initial conditions - planets} \label{ICplanets}

\begin{table*}
\caption{Abbreviations used in this study, for populations of particles that are gravitationally bound to the star cluster (column~2) and for populations of particles that have escaped from the star cluster (column~3). 
\label{tab:table}}
\begin{tabular}{lll}
\hline
Particle  & Cluster member & Escaped from cluster \\
\hline
Intermediate-mass black hole & \bh & --- \\
Star-planet system  & \bbp{} & e\bbp \\
Primordial free-floating planet & \ffp & e\ffp \\
Bound free-floating planet; a free-floating planet that was in orbit around a star at $t=0$ & \bfp & e\bfp \\
Non-host stars (including stars that lost their planet) & \nhs  & e\nhs \\
\bottomrule
\end{tabular}
\label{table:abb}
\end{table*}

We study the dynamical evolution of two populations of planets: (i) planets in orbit around a star s), and (ii) free-floating planets. Their main properties are summarised in Table~\ref{table:initialconditions}. The abbreviations used in this paper are listed in Table~\ref{table:abb}.

In each star cluster, five hundred randomly selected stars are assigned a Jupiter-mass planetary companion. Semi-major axes are drawn from a uniform logarithmic distribution in the interval $1~{\rm AU}<a<100$~AU. In Section~\ref{auchange} we describe two additional models, where we consider the intervals $a=1-10$~AU and $a=10-100$~AU. All planets are initialised on circular ($e=0$) orbits. Hereafter, we refer to such a star-planet systems as a \bbp{} when it is gravitationally bound to the star cluster, and as an e\bbp{} when it has escaped from the star cluster.

In addition, we add five hundred Jupiter-mass free-floating planets (\ffp{}s) to each star cluster. Their initial positions and velocities are drawn from distributions that are statistically identical to those of the stars, and are therefore initially in virial equilibrium with the stellar population.

Stellar encounters may disrupt planetary systems and thus generate free-floating planets. In order to obtain insight into the evolution of a population of initially-virialized free-floating planets and planets expelled from their host planetary system, we distinguish these two populations. We refer to (primordial) free-floating planets in the star cluster as \ffp{}s, and to the population of planets in the cluster that was expelled from their host systems as \bfp{}s.  When these free-floating planets have escaped from the star cluster, we refer to them as e\ffp{}s, e\bfp{}s, respectively.

Stars without a planetary companion, including those that have lost their companion at an earlier time, are referred to as \nhs{}s when their are gravitationally bound to the star cluster, and as e\nhs{} when they have escaped.
We do not describe planetary capture events, which are rare, and refer to Shu et al. (submitted) for a detailed discussion.

%%%%%%%%%%%%%%%%%%%%%%%%%%%%%%%%%%%%%%%%%%%
%%%%%%%%%%%%%%%%%%%%%%%%%%%%%%%%%%%%%%%%%%%
%%%%%%%%%%%%%%%%%%%%%%%%%%%%%%%%%%%%%%%%%%%

\subsection{Numerical method} \label{sec:style}

We use \nb{} \citep{wang2015, wang2016} to model the dynamical evolution of star clusters by direct $N$-body simulation. \nb{} is based on the earlier $N$-body codes \nbody{} \citep{Aarseth1999} and \nbodyplus{} \citep{spurzem1999}, but the main difference is its ability to use graphical processing units (GPUs). The parallelisation is achieved via \mpi{} \citep[Message Passing Interface;][]{mpi} and OpenMP on the top level, and parallel use of many GPU cores on the bottom level. The code uses the Ahmad-Cohen neighbour scheme \citep{ahmad1973}: regular forces from distant particles are parallelized using the GPU, while irregular forces from neighbour particles are computed in parallel on multi-cores on the CPU using OpenMP. The GPU implementation in \nb{} provides a significant acceleration, especially for the long-range (regular) gravitational forces. The simulation data are stored in HDF5 format \citep[see, e.g.,][]{hdf5}, which is a highly efficient storage scheme that can be used for reconstructing the dynamical properties of the star clusters with high temporal and spatial accuracy for further analysis \citep{bts}. Stellar evolution in \nb{} is modelled using the prescriptions of \cite{hurley2000, hurleypois2002, hurley2005}, for single stellar evolution \citep{hurley2013sse}, for binary stellar evolution \citep{hurley2013bse}, and for fallback and kicks for the formation of stellar remnants is modelled using the prescriptions of \cite[respectively for fallbacks and kicks,][]{hobbs2005,belczynski2002}. In recent years, LIGO data helped in the updates of \nb{} and \nbseven{} codes \citep[see, e.g.,][and references therein]{belczynski2008, banerjee2020}.

Following \cite{spurzem2009}, we model planetary systems containing only a single planet. The dynamical evolution of such star-planet binary systems can be accurately modelled using the algorithms in \nb{}, with the main difference of an upgraded code version, improved to use planetary particles and a central \bh{} simultaneously and efficiently. As the star cluster evolves, stars can escape from the cluster through tidal evaporation or dynamical ejection. Following \cite{gnbs}, we identify particles with a cluster-centric radius of $r > 2\rtidal$ as escapers, where $\rtidal \propto \mcluster^{1/3}(t)$ is the instantaneous tidal radius of the star cluster, which is evaluated in \nb{} following the prescriptions of \cite{gnbs}.

%%%%%%%%%%%%%%%%%%%%%%%%%%%
%%%%%%%%%%%%%%%%%%%%%%%%%%%
%%%%%%%%%%%%%%%%%%%%%%%%%%%

\section{Results} \label{section:results}

\subsection{Theoretical estimates \label{bhexp}}

In this section we estimate the impact of a central \bh{} presence on planetary systems in a star cluster. We quantify the relative contributions of the tidal force exerted by the \bh{} and by the nearest neighbour star at different locations in the star cluster. We obtain analytical estimates for two models: for the homogeneous spherical model and the \cite{plummer1911} model. For comparison with the simulations, we also present a brief analysis on the stability of a planetary system in the presence of an isolated \bh.

%%%%%%%%%%%%%%%%%%%%%%%%%%%
%%%%%%%%%%%%%%%%%%%%%%%%%%%
%%%%%%%%%%%%%%%%%%%%%%%%%%%

\subsubsection{Formalism} \label{sec:formalism}

Let $\mu$ be the mass of the planet, $m_\star$ the mass of the host star, and $m_n$ be the mass of a neighbouring body. Let $r_\star$ be the distance between the star and the planet, and $r_n$ be the distance between the star and the neighbouring body. The gravitational force between the star and the planet is then 
\begin{equation}
F_\star = \frac{G \mu m_\star }{ r_\star^2 } \ , 
\end{equation}
and the tidal force that is exerted by the neighbouring body on the planet is then, to first order approximation:
\begin{equation}
F_n = \frac {2 G \mu m_n r_\star }{ r_n^3 } \ .
\end{equation}
The ratio between the tidal force exerted by the neighbouring body on the planet and the force exerted by the host star on the planet is
\begin{equation} \label{eq:fequation}
f = \frac{F_n}{F_\star} = 2 \left(\frac{m_n}{m_\star}\right) \left(\frac{r_\star}{r_n}\right)^3 \ .
\end{equation}
In a star cluster that contains an \bh{}, planetary systems are affected by both the neighbouring stars and by the \bh{}. Let $M_B$ and $M_n$ be the masses of the \bh{} and the nearest neighbour star, respectively. Let $R_B$ and $R_n$ be the distances to the IMBH and to the nearest neighbour star, respectively. The ratio between the two tidal forces is then
\begin{equation} \label{eq:gequation}
g = \frac{f_B}{f_N} = 2 \frac{M_B}{M_n} \left(\frac{R_n}{R_B}\right)^3 \ . 
\end{equation}
When $g=1$, the tidal forces exerted by the \bh{} on the planet is equal to that of the neighbour star. When $g<1$ ($g>1$), the neighbour star's tidal force is stronger (weaker) than that of the \bh. When inserting $g=1$ we obtain
\begin{equation} \label{eq:fourbody}
\frac{M_n}{M_B} = 2 \left(\frac{R_n}{R_{B}}\right)^3 \ .
\end{equation}

Star clusters are evolving systems, in which a planetary system experiences rapid changes in the tidal forces it experiences. The expressions above, however, can provide rough estimates about whether the tidal force experienced by a planetary system is dominated by the \bh{} or by the neighbour stars.

If the assumption is made that the nearest neighbour star to the planetary system exerts a tidal force that is much larger than that of the more distant nearest neighbours, the expressions above can be used to make estimates for a star cluster environment. In this case, $M_n=\maverage$ can be interpreted as the average mass of the nearest neighbour, and $R_n$ as the average distance to the nearest neighbour. Let $d$ be the distance from the planetary system to the star cluster centre. For star clusters in which the black hole is in the centre of the star cluster ($d=R_B$),  Eq.~(\ref{eq:fourbody}) can be used to obtain the distance form the star cluster centre at which the tidal contribution of the \bh{} is equal to that of the nearest neighbour star, under the assumption that both $R_n$ and $M_n$ are independent of $d$:
\begin{equation}
d=R_{B} = R_{n} \left(\frac{2 M_{B}}{M_n}\right)^{1/3}  \quad .
\label{general}
\end{equation}
In the more general case of a star cluster, the properties of the nearest neighbour stars may depend on location. In these cases, solving Eq.~(\ref{eq:gequation}), with $R_B=d$, $R_n=R_n(d)$ (a radial density profile), and $M_n=M_n(d)$ (mass segregation) may result in zero, one, or multiple solutions when solving $g(d)=1$. 

%%%%%%%%%%%%%%%%%%%%%%%%%%%
%%%%%%%%%%%%%%%%%%%%%%%%%%%
%%%%%%%%%%%%%%%%%%%%%%%%%%%

\subsubsection{The homogenous spherical star cluster} \label{sec:sphere}

Consider a spherical and homogeneous star cluster of radius $\rcluster$ containing $N$ stars and an \bh{} of mass $M_B$ at its centre. Let $d$ again be the distance of a planetary system to the centre of the star cluster. 

The typical distance between the planetary system and its nearest neighbour is $R_n \approx N^{-1/3}\rcluster$ when the planetary system is inside the star cluster, and $R_n \approx d-\rcluster$ when the planetary system is outside the star cluster ($d>\rcluster$). The tidal ratio, as defined in Eq.~(\ref{eq:gequation}) is then
\begin{equation} \label{eq:g_for_spherical}
g(d) \approx
\left\{
\begin{array}{ll}
2\frac{M_B}{M_{\rm cl}}\left(\frac{\rcluster}{d}\right)^3 
& d \leq \rcluster\\
2\frac{M_B}{\maverage}\left(1-\frac{\rcluster}{d}\right)^3 
& d > \rcluster
\end{array}
\right.
\end{equation}
When considering a planetary system inside the star cluster ($d<\rcluster$), Eq.~(\ref{general}) becomes
\begin{equation}
d = R_B = \rcluster \left(\frac{2 M_{B}}{N\maverage}\right)^{1/3} 
 = \rcluster \left(\frac{2 M_{B}}{\mcluster}\right)^{1/3}  \ ,
\end{equation}
where $\mcluster$ is the total mass of the star cluster.

This allows us to estimate the region in which the tidal force of the \bh{} dominates over that of the a planetary system's neighbour stars. For a homogeneous cluster containing $N=10\,000$ stars and $\rcluster=1$~pc, we obtain
\begin{equation}
R_{\rm bh} = 0.069 \left(\frac{M_B}{M_\odot} \right)^{1/3} \quad {\rm pc} \ . 
\end{equation}
Here we have adopted the average stellar mass $\maverage \approx 0.6~\msun$ of the \cite{kroupa2001} IMF. 

For a star cluster with a central \bh{} of mass of $\mbh=100\msun$, the radius of the sphere in which the \bh{} dominates the tidal evolution of the planets is $R_{\rm bh}=0.320$~pc. This means that roughly 32\% of the stars in the star cluster fall in this category. For star clusters with an \bh{} mass of $M_B=200\msun$ the value increases to $R_{\rm bh} = 0.403$~pc, representing roughly 40\% of the stars in the cluster.

%%%%%%%%%%%%%%%%%%%%%%%%%%%
%%%%%%%%%%%%%%%%%%%%%%%%%%%
%%%%%%%%%%%%%%%%%%%%%%%%%%%

\subsubsection{The Plummer model} \label{sec:plummer}

The \cite{plummer1911} model provides a more accurate description of star clusters than that described in Section~\ref{sec:sphere}. 
The mass density profile for the Plummer model is 
\begin{equation} \label{eq:massdensity}
\rho_m (d) = \frac{3\mcluster}{4 \pi \alpha^3} 
\left(1+ \frac{d^2}{\alpha^2}\right)^{-5/2} 
\end{equation}
\citep[see, e.g.,][]{heggie2003}, where $\alpha = 3 \pi \rvir/16$ is the scale radius, and $\mcluster$ the total mass of the star cluster (excluding the \bh). The number density of stars can be obtained by dividing Eq.~(\ref{eq:massdensity}) by the average stellar mass:
$ \rho_\star (d) = \rho_m (d)/\maverage$. 

At a distance $d$ from the cluster centre, the average distance between a planetary system and its nearest neighbour star is then 
\begin{equation} 
R_n(d)=\rho_\star(d)^{-1/3} =  
\left(\frac{4\pi\alpha^3}{3N}\right)^{1/3} 
\left(1+ \frac{d^2}{\alpha^2}\right)^{5/6}
\ ,
\end{equation}
where we have used $\mcluster = \maverage N$. Substituting the above expressions into Eq.~(\ref{eq:gequation}), and assuming that the \bh{} is located at the centre of the star cluster, we obtain 
\begin{equation} \label{eq:gplummer}
g(d) = 
\frac{8\pi}{3}
\frac{\mbh}{\mcluster}
\left(\frac{\alpha}{d}\right)^3
\left(1+ \frac{d^2}{\alpha^2}\right)^{5/2}
\ .
\end{equation}
Near the cluster centre (i.e., near the \bh{}), where $d \ll \alpha$, the above expression reduces to 
\begin{equation} 
g(d) = 
\frac{8\pi}{3}
\frac{\mbh}{M}
\left(\frac{\alpha}{d}\right)^3 
\gg 1 \ .
\end{equation}
In the region surrounding the \bh{}, the influence of the \bh{} is much stronger than that of the nearest neighbour star. Far away from the star cluster $(d \gg \alpha)$, Eq.~(\ref{eq:gplummer}) reduces to 
\begin{equation} \label{eq:gapprox1}
g(d) = 
\frac{8\pi}{3}
\frac{\mbh}{M}
\left(\frac{d}{\alpha}\right)^2
\gg 1 \ ,
\end{equation}
which tells us that the tidal force of the \bh{} on the planetary system is much larger than that of the {\em nearest} neighbour.

The minimum of $g(d)$ occurs at $d_{\rm min}=(3/2)^{1/2}\alpha \approx 1.22\alpha$. Note that the location of this minimum is independent of the number of stars in the cluster, the mass of the black hole, and the average stellar mass. At this minimum, the \bh{} has the smallest relative contribution to the tidal force (with respect to that of the nearest neighbours) experienced by a planetary system. At this location the tidal ratio is 
\begin{equation} \label{eq:gapprox2}
g(d_{\rm min}) \approx 29.2 \ 
\frac{\mbh}{\mcluster} \ .
\end{equation}
The tidal ratio $g(d_{\rm min})$ is smaller than unity when the \bh{} mass is less than 3.4\% of the total stellar mass of the star cluster. Note that, when the $\mbh > 0.034\mcluster$, then $g(d)>{\bf 1}$ throughout the entire cluster. In other words, the tidal force exerted by an \bh{} tends to dominate the evolution of planetary systems at all locations in the star cluster. 

The contributions of neighbouring stars and the \bh{} to the tidal force on a planetary system when $g(d)=1$. Eq.~(\ref{eq:gplummer}) has no solution of $g(d)=1$ when $\mbh>0.034\mcluster$, has one solution when $\mbh=0.034\mcluster$, and has two solutions when $\mbh<0.034\mcluster$. The latter case is most realistic for both open clusters and globular clusters.

%%%%%%%%%%%%%%%%%%%%%%%%%%%
%%%%%%%%%%%%%%%%%%%%%%%%%%%
%%%%%%%%%%%%%%%%%%%%%%%%%%%

\subsubsection{Applications} \label{sec:numericalestimates}

\begin{figure}
\centering
\includegraphics[width=0.4\textwidth,height=!]{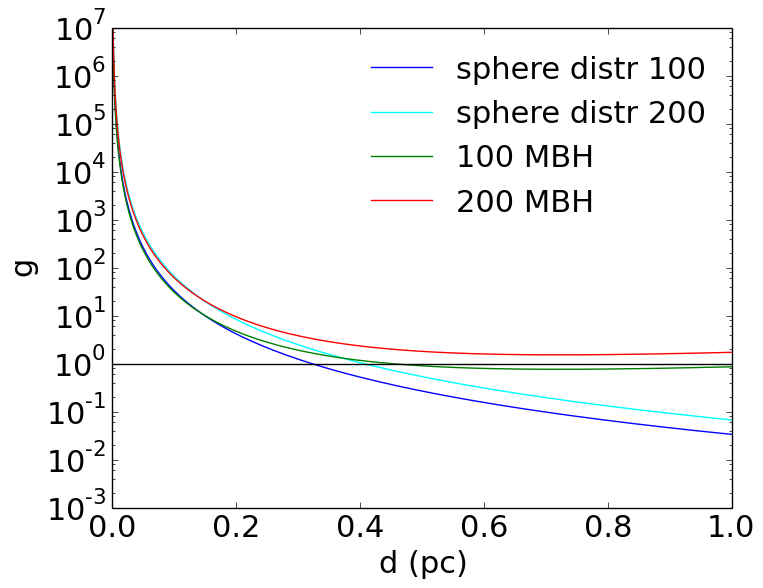} \\
\includegraphics[width=0.4\textwidth,height=!]{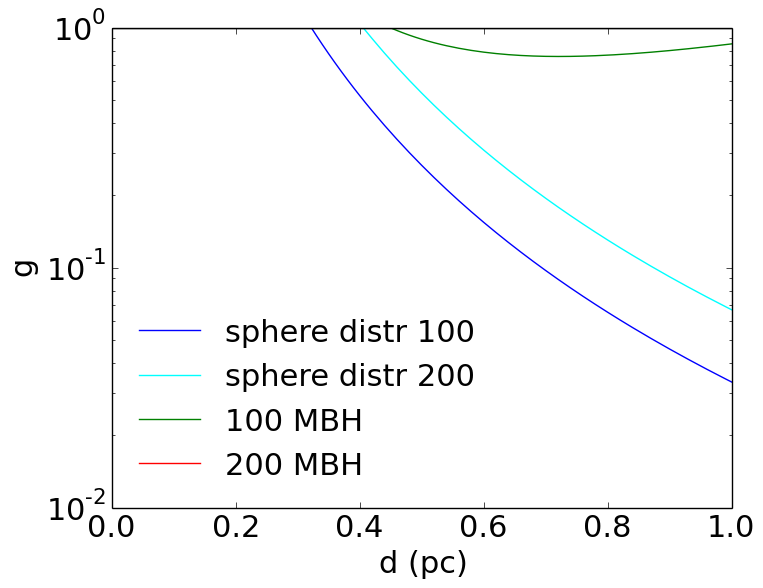}
\caption{The tidal influence ratio, $g(d)$, for the spherical homogeneous star cluster (light-blue and dark-blue curves) and for the Plummer models (green and blue curves), for $\mbh=100~\msun$ and $\mbh=200~\msun$ ({\em top panel}). The black horizontal line at $g(d)=1$ indicates the transition between tidal domination by the \bh{} and tidal domination by neighbour stars. The region with $g<1$ is shown separately in the bottom panel. } 
\label{figure:fig0}
\end{figure}

\begin{figure}
\includegraphics[width=0.5\textwidth,height=!]{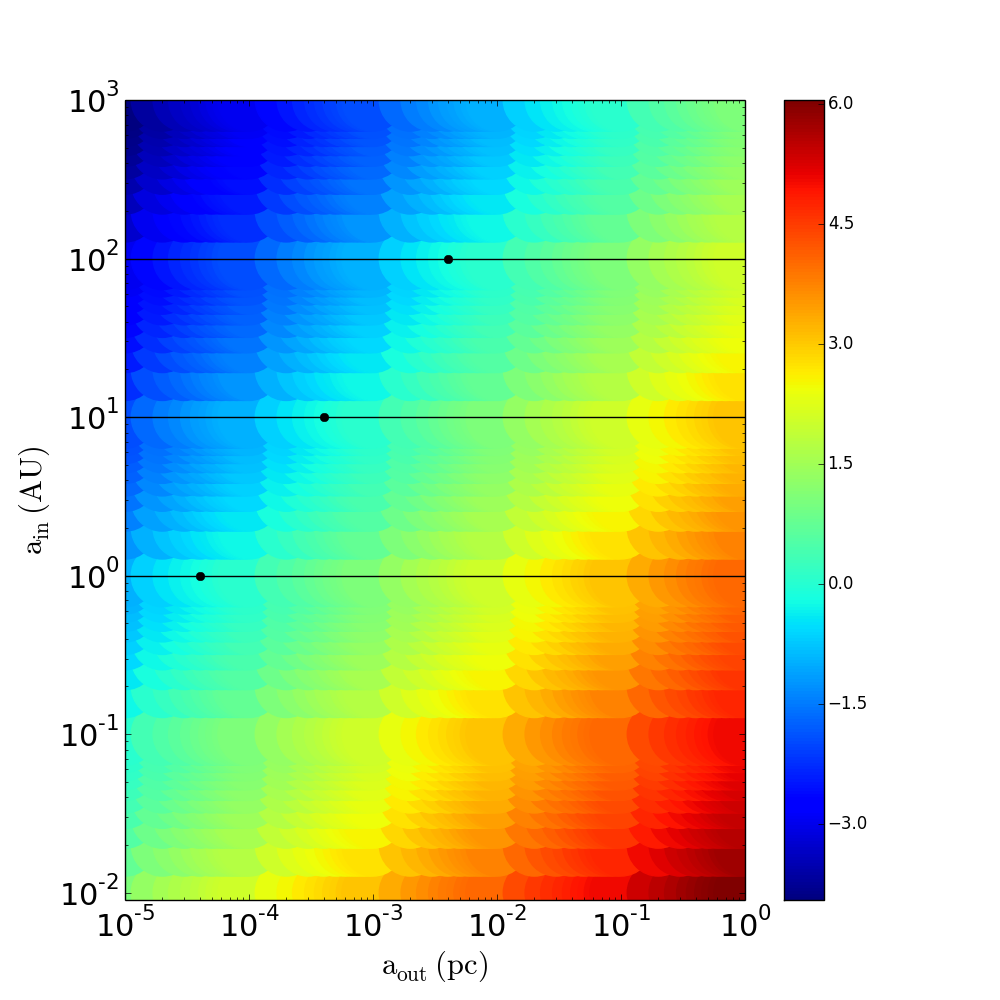}
\caption{Stability map of planetary systems, with a semi-major axis ($a_{\rm in}$) at a distance from an \bh{} ($a_{\rm out}$). The three horizontal lines represent the reference values for the semi-major axes used in our study, and the black dots are where the instability starts for each of them. The colours indicate the values of $\log (p_{\rm out}/p_{\rm crit})$. The blue regions are most unstable, while the red regions are most stable. \label{figure:stabilitymap}  }
\end{figure}

The derivations above allow us to make compare the tidal influence of \bh{}s in star clusters numerically. In these examples we consider (i) homogeneous spheres with $\rcluster=1$~pc, and (ii) Plummer models with $\rvir=1$~pc (i.e., $\alpha = 0.589$~pc). Both clusters have a total stellar mass of $\mcluster = 5950\msun$ (excluding the \bh{}). In both models, we calculate the tidal ratio $g(d)$ for star clusters with a central \bh{} masses of $\mbh=100\msun$ and $\mbh=200\msun$.

The top panel Figure~\ref{figure:fig0} shows a comparison between the ratios $g(d)$ for the four different models.  In all models $g(d) \gg 1$ in the central regions of the cluster.  In the cores of the star cluster ($d\la 0.1$~pc), the tidal force exerted by the \bh{} is orders or magnitude larger than that of the neighbour stars For the spherical homogeneous star clusters, \bh{}s tend not to dominate the tidal force experienced by planetary systems in the outer regions of the star cluster. For the Plummer models, the \bh{} plays an important role throughout the entire star cluster. Note that for $\mbh \ga 200~\msun$, $g(d)>1$ throughout the entire star cluster.

Even though in some regions of the star cluster the tidal force exerted by a central \bh{} on a planetary system is larger, and sometimes even many orders of magnitude larger, than that of the typical neighbour star, the \bh{} itself may not be the direct cause of the disruption of planetary systems. It can, however, induce orbital changes that may result in the decay of multi-planet systems, or that may facilitate disruption by neighbour stars. We discuss the stability criteria of a planetary system orbiting an isolated \bh{}. In this case, the configuration can be treated as a hierarchical triple system. We define the star-planet system as the inner binary, while the planetary system orbiting the \bh{} as the outer binary. The quantities $a_{\rm in}$, $a_{\rm out}$ represent the semi-major axes of the inner and outer orbits, $e_{\rm in}$ and $e_{\rm out}$  the inner and outer eccentricities, and $q_{\rm in} = m_{\rm p}/m_{\rm star}$ and $q_{\rm out} = \mbh/(m_{\rm star}+m_{\rm p})$ the inner and outer mass ratios. The periastron separation of the outer orbit is $p_{\rm out}=a_{\rm out}(1-e_{\rm out})$. For simplicity, we only consider co-planar orbits, i.e., systems with inclinations $i_{\rm in}=0^\circ$ and $i_{\rm out}=0^\circ$. \cite{mardling2001} numerically obtained stability criteria for hierarchical three-body systems, and find the empirical formula  
\begin{equation}
p_{\rm crit} = 2.8 \, a_{\rm in} \frac{1-e_{\rm out}}{1+e_{\rm in}} \left[(1+q_{\rm out}) \frac{1+e_{\rm out}}{(1-e_{\rm out})^{\frac{1}{2}}} \right]^{0.4} \ ,
\end{equation}
where systems with $p_{\rm out} > p_{\rm crit}$ are considered stable.

Figure~\ref{figure:stabilitymap} shows a map of the ratio $p_{\rm out}/p_{\rm crit}$ as a function of $a_{\rm out}$ and $a_{\rm in}$, for the circular, co-planar case with $\mbh=100~\msun$ \bh{}, a Solar-mass star and a Jupiter-mass planet. The blue regions in Figure~\ref{figure:stabilitymap}  indicate configurations that are unstable (i.e., those that result in disruption of the planetary system). Note that these are configurations that are unlikely to occur in realistic systems. The substantial increase in disruptions of planetary systems in the presence of an \bh{} is therefore an indirect, rather than a direct consequence of the \bh{}.

The impact of an encounter on a planetary system (in terms of its ability to cause permanent changes) is determined by (i) the kinematic properties of the encountering star, and (ii) the properties of the planetary system. An encounter with a stellar neighbour can be characterised through its impulsiveness, adiabaticity, and the hyperbolic eccentricity  \citep[e.g.,][and references therein]{flammini2019}. A planetary system experiences many of such encounters, depending on the local stellar density. While there is only one \bh{} in the star cluster, its interaction with the planetary system is permanent. The interaction with the \bh{} can thus be considered as an encounter that lasts for billions of years. The impact of this interaction can be substantial, especially when $g(d)>1$.

We remind the reader that the derivations above has several approximations, uses averaged values for stellar masses and stellar separations. It is also important to note that the presented values represent the initial conditions, and do not take into account the dynamical evolution of the star cluster.  However, these results provide useful insight into the importance of \bh{}s on the dynamical evolution of planetary systems star clusters. 

%%%%%%%%%%%%%%%%%%%%%%%%%%%%%%%%%%%%%%%%%%
%%%%%%%%%%%%%%%%%%%%%%%%%%%%%%%%%%%%%%%%%%
%%%%%%%%%%%%%%%%%%%%%%%%%%%%%%%%%%%%%%%%%%

\subsection{Star cluster evolution} \label{sec:starclusterevolution}

Although the focus of this study is the effect of an \bh{} on the evolution of planetary systems in star clusters, it is important to first consider how an \bh{} considers the star cluster itself, as the latter can have a considerable indirect effect on how planetary systems evolve these environments.

%%%%%%%%%%%%%%%%%%%%%%%%%%%%%%%%%%
%%%%%%%%%%%%%%%%%%%%%%%%%%%%%%%%%%
%%%%%%%%%%%%%%%%%%%%%%%%%%%%%%%%%%

\subsubsection{Evolutionary timescales} \label{sec:timescales}

To understand how the environment of planetary systems in star clusters is shaped by interactions between the stellar population and the \bh{} it is useful to consider the relevant timescales at which these changes occur. Stars cross the star cluster at a timescale that is comparable to the crossing time,
\begin{equation}
t_{\rm cr} = \sqrt{\frac{2 \halfmass^{3}}{G \mcluster }}
\end{equation}
\citep[e.g.,][]{spitzer1987, lamers2005,binney2008}. Here $\halfmass$ is the half-mass radius, $G$ is the gravitational constant, and $M_{\rm cl}$ the total mass of the star cluster. 
The relaxation time is defined by the time over which, for most stars in the cluster, the effect of stellar encounters on the velocity of stars becomes comparable to the star's initial velocity:
\begin{equation}
\relaxationtime = \frac{0.138 N_s}{\ln \Lambda} \sqrt{\frac{\halfmass^3}{G \mcluster}}
\end{equation}
\citep[e.g.,][]{spitzer1987, khalisi2007}, where $\Lambda$ depends on the stellar density distribution of the cluster \citep[as in][]{spitzer1987}: the corresponding value is $\Lambda = 0.11 N_s$ \citep{giersz1994}.

The energy equipartition timescale (often referred to as the mass segregation timescale) describes how long it takes for different stellar populations within the star cluster to exchange energy \cite[see, e.g.,][]{spurzem1995, mouri2002}:
\begin{equation}
t_{\rm ms} = \frac{m_s}{\maverage} \relaxationtime
\end{equation}
\citep{khalisi2007}, where $m_s$ is the stellar mass under consideration and $\maverage$ is the average stellar mass. 
The mass segregation mechanism appears to be quenched in globular clusters that contain a central \bh{} \citep{gill2008}.

The core collapse timescale depends on both the star cluster's initial density profile and on the properties of the binary population  \citep[e.g.,][]{hurley2007b}.
\cite{heggie2003} define the core collapse timescale as 
\begin{equation}
t_{\rm cc} \approx \frac{v^3}{\nu G^2 \maverage^2 n \log \Lambda } \ ,
\label{eq:cc}
\end{equation}
where $\nu$ is a parameter related to the relaxation timescale and varies in time, $v$ is the typical stellar speed in the core, $\maverage$ the typical stellar mass in the core and $n$ the density number in the core. For the Plummer model, $t_{\rm cc} \approx 15\,t_{\rm rh,core}$, \citep[see, e.g.,][]{pz2010}.

%%%%%%%%%%%%%%%%%%%%%%%%%%%%%%%%%%%%%%%
%%%%%%%%%%%%%%%%%%%%%%%%%%%%%%%%%%%%%%%
%%%%%%%%%%%%%%%%%%%%%%%%%%%%%%%%%%%%%%%

\subsubsection{Global properties} \label{sec:coreradius}

\begin{figure}
\centering
\includegraphics[width=0.4\textwidth,height=!]{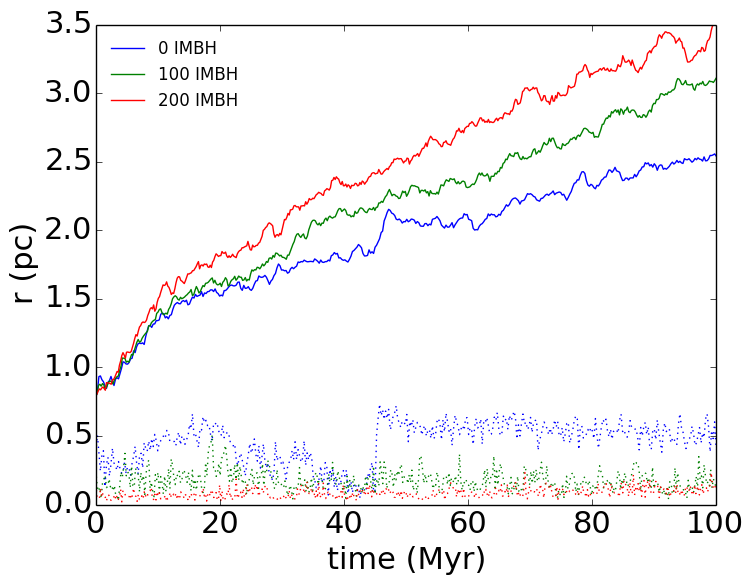}
\caption{The evolution half-mass radii ({\em solid curves}) and core radii ({\em dotted curves}), for models C50M00 ({\em blue}), model C50M100 ({\em green}), and model C50M200 ({\em red}); see also Table~\ref{table:initialconditions}. }
\label{figure:clusterevolution}
\end{figure}

\begin{figure}
\centering
\includegraphics[width=0.4\textwidth,height=!]{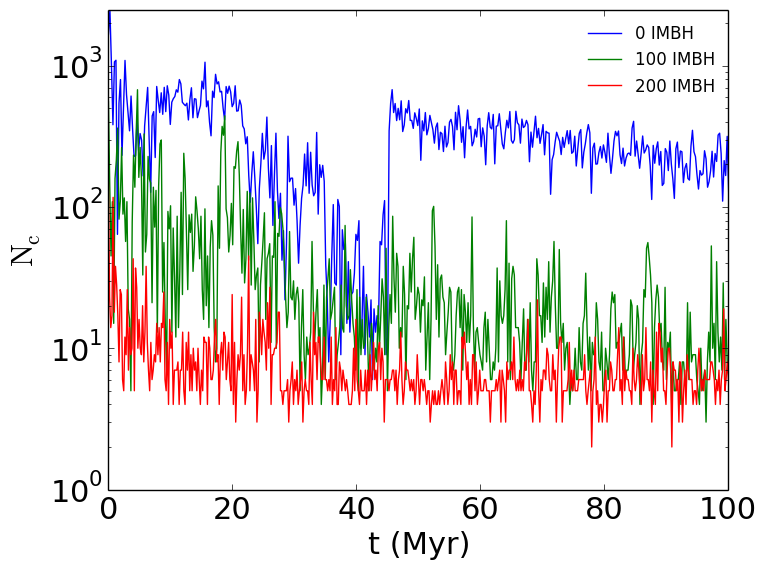}\\
\includegraphics[width=0.4\textwidth,height=!]{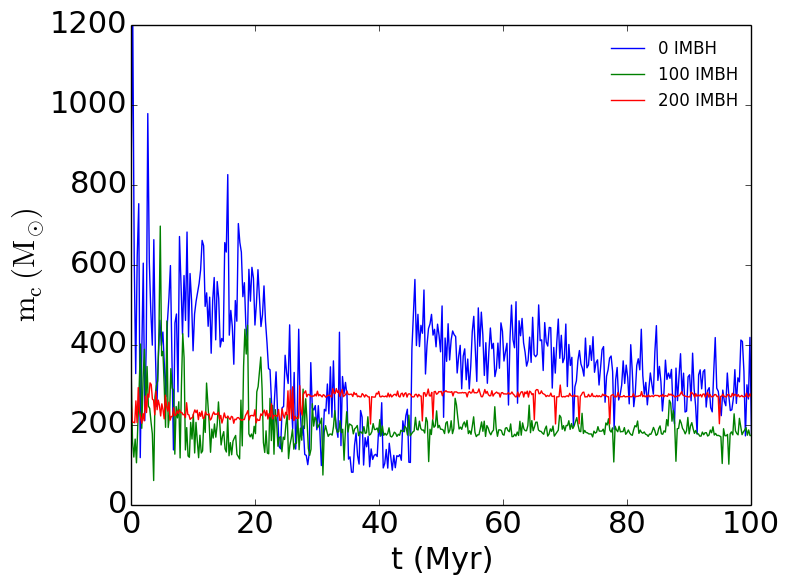}
\caption{Evolution of the number of stars in the cluster core ({\em top}) and the enclosed mass in the cluster core ({\em bottom}), for the three reference models. \label{figure:chnandm}  }
\end{figure}

\begin{figure}
\centering
\includegraphics[width=0.4\textwidth,height=!]{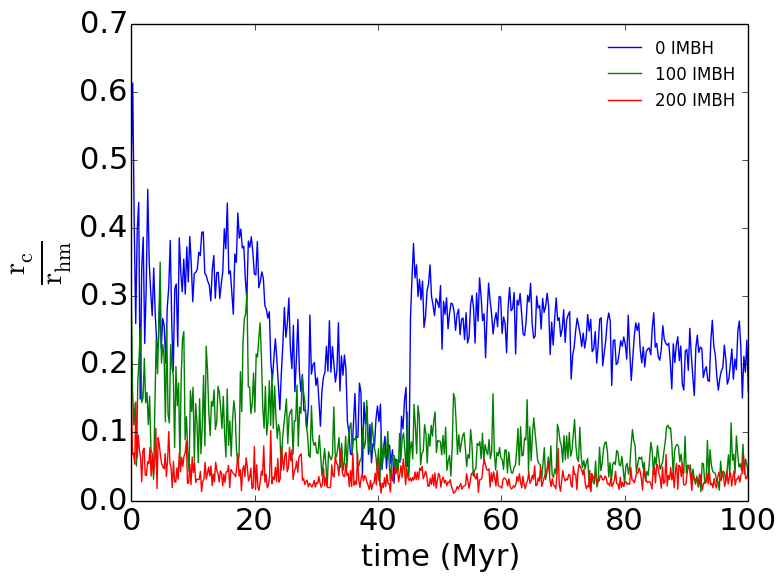}
\caption{Evolution of the ratio between the core radius and half-mass radius for the three reference models (c.f. Figure~\ref{figure:clusterevolution}). }
\label{figure:rchmrtime} 
\end{figure}

The presence of a central \bh{} affects the global properties of a star cluster that can be observed in multiple ways: (i) the evolution of the core radius, (ii) the speed at which the energy equipartition process operates, and (iii) the delay, or prevention, of core collapse. Moreover, the kinematics of all constituent populations of the star cluster is directly and/or indirectly affected by the presence of the \bh{}.

In observational astronomy, the core radius of a star cluster is often defined as the radius at which the surface brightness drops to one-half its central value \citep[e.g.,][]{wilkinson2003}. Different definitions are  used for theoretical and computational studies, where the core radius is normally derived from the three-dimensional mass density profile of the star cluster \citep{casertano1985}. \cite{aarseth2001} introduced the density radius as a measurement of the size of a star cluster in the \nbody{} code. The stellar density at the location of each star is estimated from the mass of the sphere containing the six nearest neighbour stars, while the density centre is defined as the density-weighted mean position of the stars, following the prescription of \cite{casertano1985}.

The impact of \bh{}s on the evolution of our modelled star clusters is shown in Figure~\ref{figure:clusterevolution}. The \bh{} makes a considerable contribution to the mass profile, and consequently the Lagrangian radii, in the core region. 
The interaction of the \bh{} with surrounding stars causes an accelerated expansion of the star cluster. During the first few million years, the half-mass radius grows faster for the models containing an \bh{}. The core particles escape in a shorter time due to the \bh{}'s gravitational kicks, and tends to enlarge the half-mass radius when $t \la \relaxationtime$. The star cluster relaxes and the $\halfmass$ evolves similarly in all the three models at subsequent times.
The evolution of the core radius is more complex. In model C50M000, the core shrinks at $t \approx \relaxationtime$ and subsequently grows again at $t \approx 2.5\relaxationtime$, as a result of mass segregation, followed by a core collapse \citep[e.g.,][]{khalisi2007}. 
The core of model C50M000 enlarges after core collapse has occurred, and also ejects stars towards the outer regions of the cluster, reducing the total number of stars in the core.
Although the core evolution appears steady during the first 100~Myr for the models containing an \bh{}, the number of particles varies in the core until $\sim 40$~Myr, leaving a small number of particle members for the remaining 60~Myr.

Figure~\ref{figure:chnandm} provides us insights on the evolution of the star cluster core. In model C50M00 (which does not contain an \bh{}), the number of stars in the core, as well as the mass of the core, decreases substantially until the core collapses at $t \approx 40$~Myr, and then rises again. 
In the models that contain a central \bh{}, the number of stars in the core is substantially reduced, reaching a minimum of 0.1\% of the total number of stars in model C50M100, and 0.03\% for model C50M200. 
In general, the number of stars in the cluster core at any given time is smaller when a cluster has a more massive \bh. The presence of an \bh{}, however, appears to prevent the cluster core from collapsing, and a core with a more or less constant stellar density and mass density is maintained for long periods of time.
In star cluster without an \bh{}, dynamical interactions in the cluster core occur much faster in the timespan surrounding the core collapse. This state of high stellar density does not occur when a central \bh{} is present, because of the quenched mass segregation. This phenomenon is described in \cite{gill2008}, who carried out a numerical investigation of the quenching mechanism in globular clusters, in which the \bh{} prevents core collapse. 
Figure~\ref{figure:rchmrtime} shows the evolution of the ratio between the core radius and the half-mass radius for the three reference models. During the observational analysis of star clusters, the ratio between the core radius and the half-mass radius, combined with the ratios of the enclosed number of stars and the enclosed total mass in the cluster core, relative to those of the half-mass radius, can be used to identify when mass segregation has been quenched \citep{gill2008}. Note that we have included stellar evolution, which may cause subtle differences between our results and those of \cite{gill2008}.

%%%%%%%%%%%%%%%%%%%%%%%%%%%%%%%%%%%%%
%%%%%%%%%%%%%%%%%%%%%%%%%%%%%%%%%%%%%
%%%%%%%%%%%%%%%%%%%%%%%%%%%%%%%%%%%%%

\subsubsection{Escaping stars and planets} \label{sec:escaperates}

\begin{figure*}
\centering
\begin{tabular}{ccc}
  \includegraphics[width=0.32\textwidth,height=!]{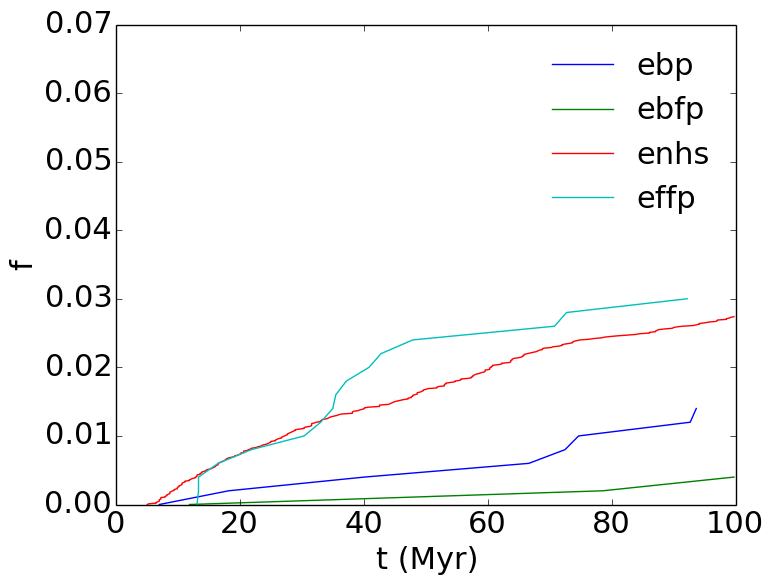} &
  \includegraphics[width=0.32\textwidth,height=!]{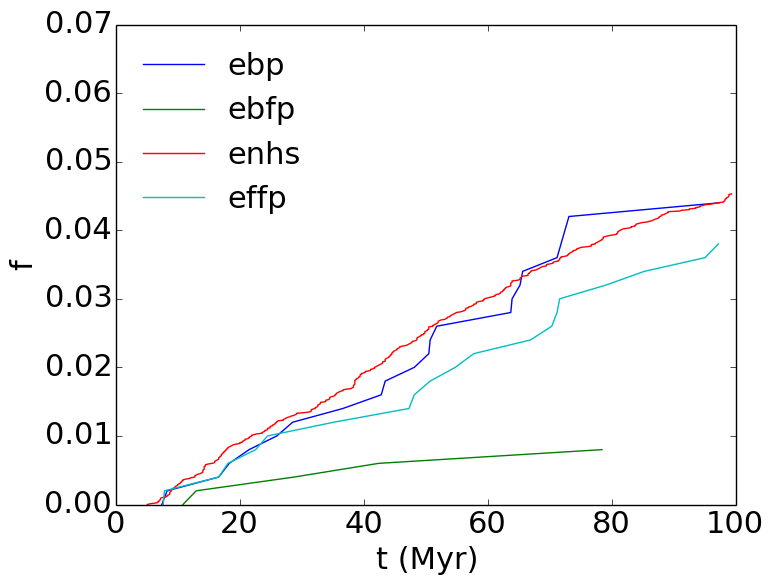}  &
  \includegraphics[width=0.32\textwidth,height=!]{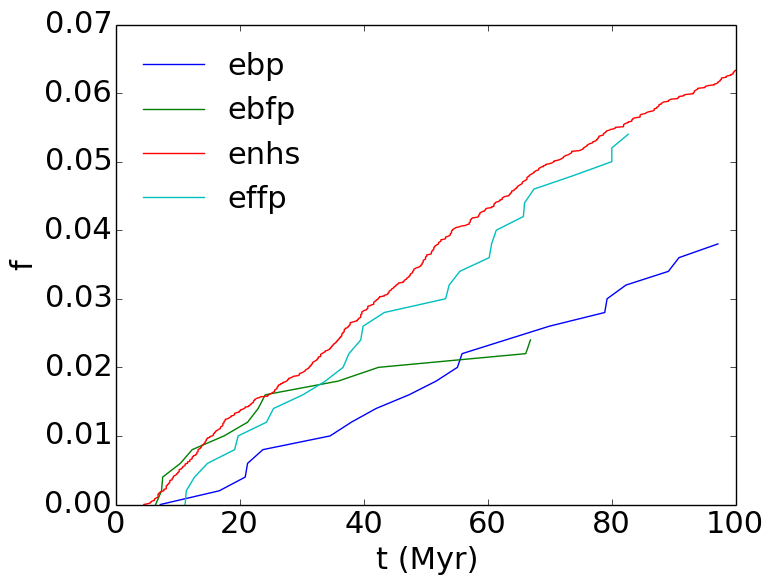}  \\
  \includegraphics[width=0.32\textwidth,height=!]{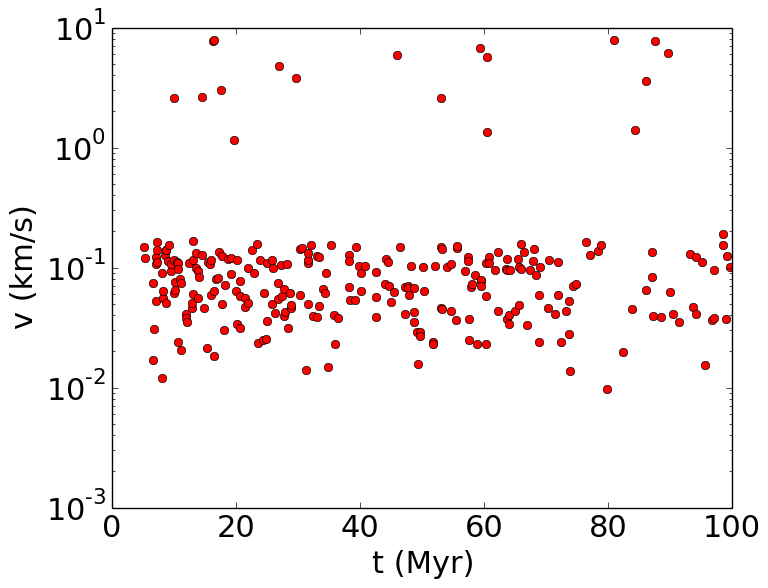} &
  \includegraphics[width=0.32\textwidth,height=!]{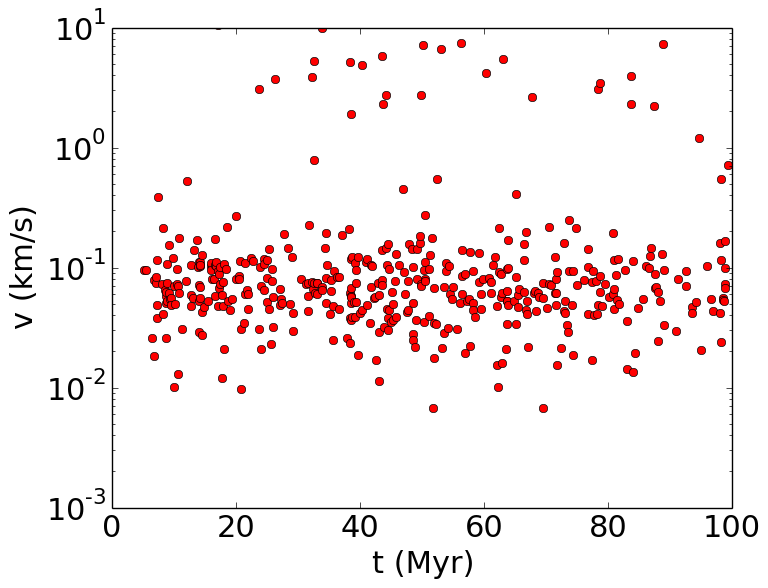}  &
  \includegraphics[width=0.32\textwidth,height=!]{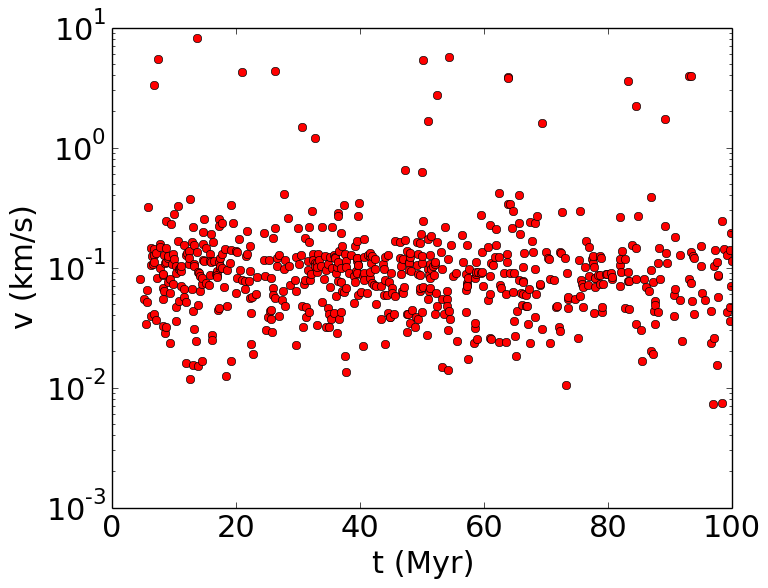}  \\
  \includegraphics[width=0.32\textwidth,height=!]{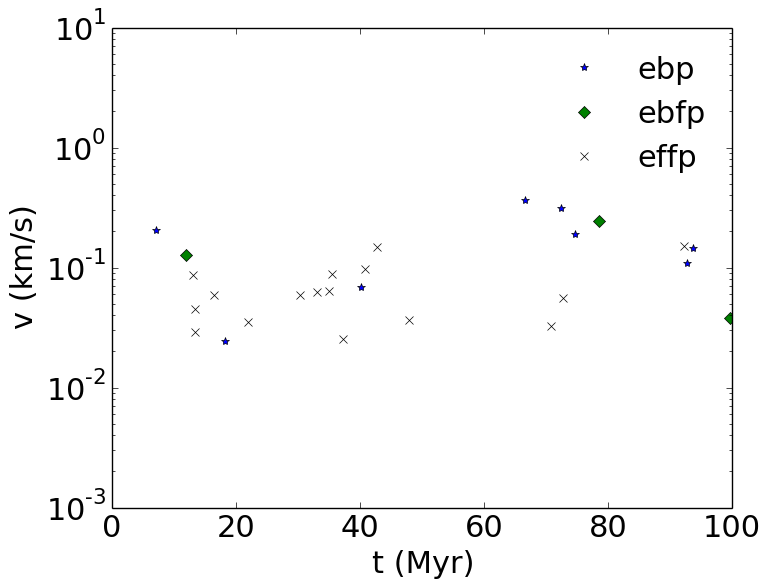} &
  \includegraphics[width=0.32\textwidth,height=!]{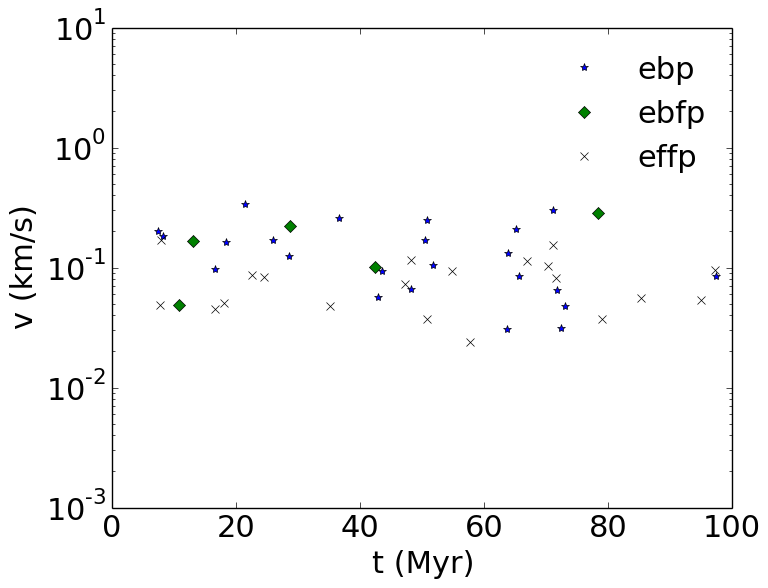}  &
  \includegraphics[width=0.32\textwidth,height=!]{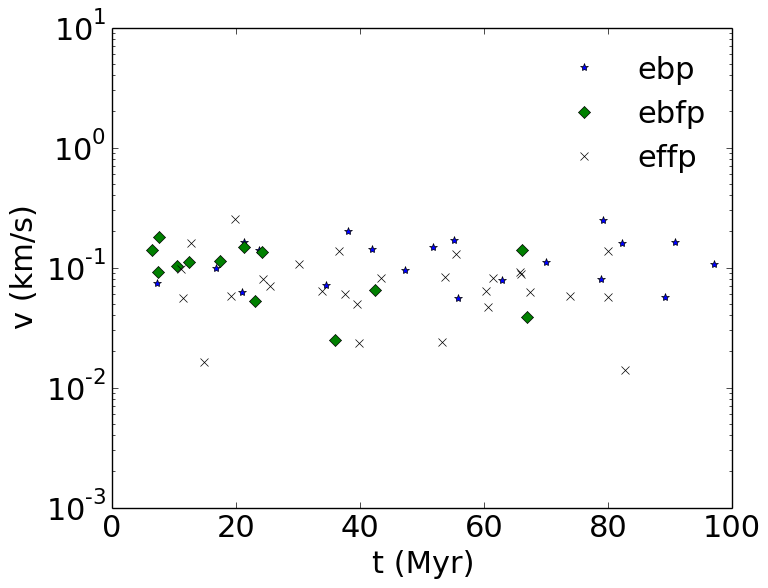}  \\
\end{tabular}
\caption{Properties of the escaping particles for models C50M000 ({\em left}), C50M100 ({\em middle}), and C50M200 ({\em right}). The top row shows the 
cumulative distributions of the fraction of planetary systems, stars, and planets escaping from the star clusters (e\bbp/\bbp$_0$, \bfp/\bbp$_0$, e\bfp/\bbp$_0$, and e\ffp/\ffp$_0$; see Table~\ref{table:abb}). The middle and bottom rows show the velocity-at-infinity versus escape time for the escaping stars and for the escaping stars, respectively. }
\label{figure:escapers1}
\end{figure*}

\begin{table*}
\caption{The fraction of particles that have escaped at $t=100$~Myr, for different star clusters. The prefix "av" indicates the averaged results for the ensemble of simulations.}
\label{tab:table}
\begin{tabular}{lccccc}
\hline
Model ID  & e\bbp/\bbp$_0$ & \bfp/\bbp$_0$ & e\bfp/\bbp$_0$ & e\nhs/$\nstars$$_0$ & e\ffp/\ffp$_0$  \\
\hline
C50M000 & $1.60 \pm 0.56 \ \%$  & $\ 8.80 \pm 1.27 \ \%$ & $0.60 \pm 0.34 \ \%$ & $2.78 \pm 0.16 \ \%$ & $3.20 \pm 0.79 \ \%$   \\
C50M100 & $4.60 \pm 0.94 \ \%$ & $10.80 \pm 1.39 \ \%$ & $1.00 \pm 0.45 \ \%$ & $4.58 \pm 0.21 \ \%$ & $4.00 \pm 0.88 \ \%$  \\
C50M200 & $4.00 \pm 0.88 \ \%$ & $10.20 \pm 1.35 \ \%$ & $2.60 \pm 0.71 \ \%$ & $6.42 \pm 0.24 \ \%$ & $5.60 \pm 1.03 \ \%$    \\
avC50M000 & $1.82 \pm 0.04 \ \%$ & $\ 9.94 \pm 0.04 \ \%$ &   $1.30 \pm 0.05 \ \%$ &  $2.72 \pm 0.02 \ \%$ & $2.24 \pm 0.02 \ \%$ \\
avC50M100 & $3.36 \pm 0.04 \ \%$ & $11.68 \pm 0.04 \ \%$&  $1.70 \pm 0.04 \ \%$ & $4.01 \pm 0.02 \ \%$ & $3.60 \pm 0.03 \ \%$  \\
avC50M200 & $3.64 \pm 0.03 \ \%$ & $11.22 \pm 0.04 \ \%$ & $2.22 \pm 0.05 \ \%$ & $5.57 \pm 0.02 \ \%$ & $5.52 \pm 0.04 \ \%$    \\
\bottomrule
\end{tabular}
\label{table:escstars}
\end{table*}

Here we analyse the escape rates of different populations in star clusters, with an emphasis on how they relate to the \bh{}. Statistics on the escape fractions at $t=100$~Myr are listed in Table~\ref{table:escstars} for the different constituent populations of the star cluster. The gravitational potential evolves over time as a consequence of two-body relaxation, stellar evolution, and mass loss due to escaping stars. 
A central \bh{} affects the gravitational potential of the star cluster directly (due to its presence in the centre of the star cluster), as well as indirectly (due to its effect on the dynamics of the stellar population, including quenching of mass segregation). This enhances both the ejection rates of stars and \ffp{}s. The kinematics of the \ffp{}s is primarily determined by the global properties of the star cluster's gravitational field \citep[e.g.,][]{wangkouwenhoven2015}.

As compared to the model without an \bh{} (model C50M000) at $t=100$~Myr, the ejected fraction of e\ffp{}s is 60\% higher in model C50M100 and 146\% higher in model C50M200. The fractions of ejected stars is 48\% and 107\% higher in models C50M100 and C50M200, respectively. At the end of the simulations ($t=100$~Myr), the number of e\bfp{}s is somewhat larger in star clusters containing an \bh{}, by 17\% and 13\%, for models C50M100 and C50M200, respectively. In latter models, a larger fraction of stars tends to migrate out of the star cluster centre, decreasing the stellar density and hence, also the close encounter rate. This effect is more prominent for clusters with higher-mass \bh{}s. Most of the encounters involving a notable interaction with the \bh{} occur before the central region expands.
The fraction of \bbp{}s ejected in model C50M000 is comparable to that of the models of \cite{flammini2019}. The average number of ejected \bbp{}s at $t=100$~Myr is 83\% and 100\% higher, for models C50M100 and C50M200, respectively, as compared to model C50M000.

The distribution of escape times is shown in the top panels of Figure~\ref{figure:escapers1}. The escape rate of all populations is higher when a more massive \bh{} is present.  The escapes rate of the different populations of particles is roughly constant with time
For model C50M000, half of the stars that are ejected during the simulation ($t<100$~Myr) are expelled from the cluster within $\sim 50$~Myr. For models C50M100 and C50M200, these timescales are $\sim 35$~Myr and $\sim 20$~Myr for C50M200, respectively. The \bh{} ejects most stars during the first few million years, resulting in a larger total number of ejections. High-mass stars tend to rapidly migrate to the cluster centre when no \bh{} is present. When an \bh{} is present, on the other hand, mass segregation is quenched, and many higher-mass stars are bounced out of the cluster core, the ejection rate may be somewhat higher.

Half of the e\ffp{}s are produced within $\sim 40$~Myr for model C50M000, while the corresponding timescales are 
45~Myr and $\sim 30$~Myr for models C50M100 and C50M200.
The ejection of \ffp{}s from the cluster is slowed down when an \bh{} is present in the star cluster. The ejection rate of \ffp{}s directly related to the evolution of the gravitational potential and the strength of the tidal field. However, the \bh{}, both directly and indirectly, remains the major catalyst for the ejection of both stars and planet s  from the cluster. Unlike model C50M000, the production rate of e\ffp{}s is roughly constant, for models C50M100 and C50M200.

The middle and bottom panels in Figure~\ref{figure:escapers1} show the escape speeds of stars and planets, respectively, as a function of the time at which each particle is identified as an escaper. The high-velocity single stars are ejected from the cluster core  following strong scattering events with other stars and/of the \bh{} \citep[e.g.,][]{gvaramadze2009},  while low-velocity stellar and planetary escapers evaporate from the star cluster outskirts. As the cluster expands and gradually fills its Roche lobe, the tidal field gradually strips off stars from its outskirts. 
The number of stars that has escaped from the star cluster at $t=\relaxationtime$ is higher in models containing a central \bh{}. At this time, the fraction of stars that have escaped is $\sim$ 1.3\%, 1.5\% and 2.0\% of models C50M000, C50M100 and C50M200, respectively. 
Most stars and planets escape with speeds between 0.01~\kms{} and 10~\kms. As both stellar evolution and stellar escapers reduce the total mass of the star cluster, the gravitational potential is reduced and, consequently, the typical escape velocity also decreases slightly over time, while the stars passing through the core have their orbit augmented by the \bh.

\subsubsection{\bh{} kinematics} \label{sec:bhkinematics}

\begin{figure*}
\centering
\begin{tabular}{ccc}
\includegraphics[width=0.32\textwidth,height=!]{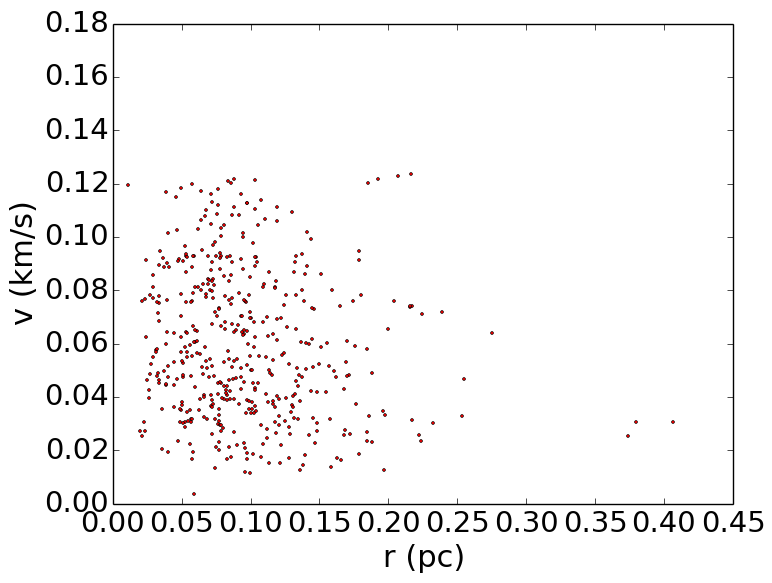} &
\includegraphics[width=0.32\textwidth,height=!]{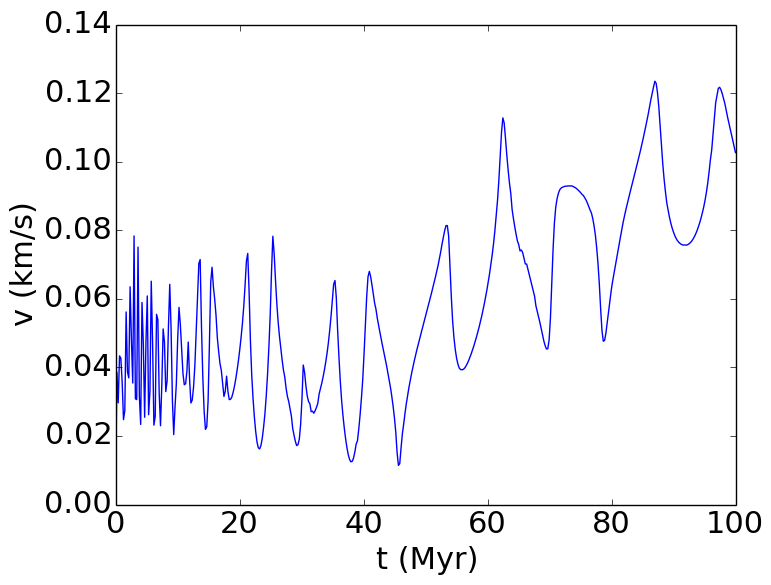} &
\includegraphics[width=0.32\textwidth,height=!]{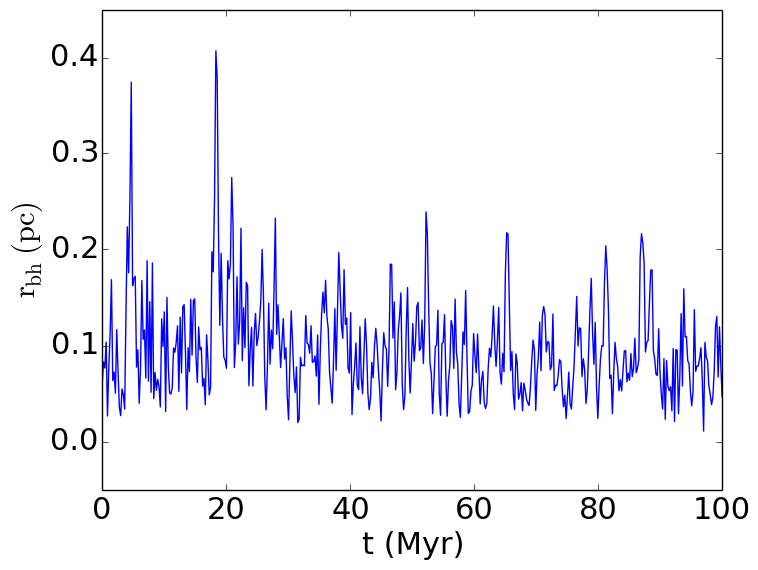}\\
\includegraphics[width=0.32\textwidth,height=!]{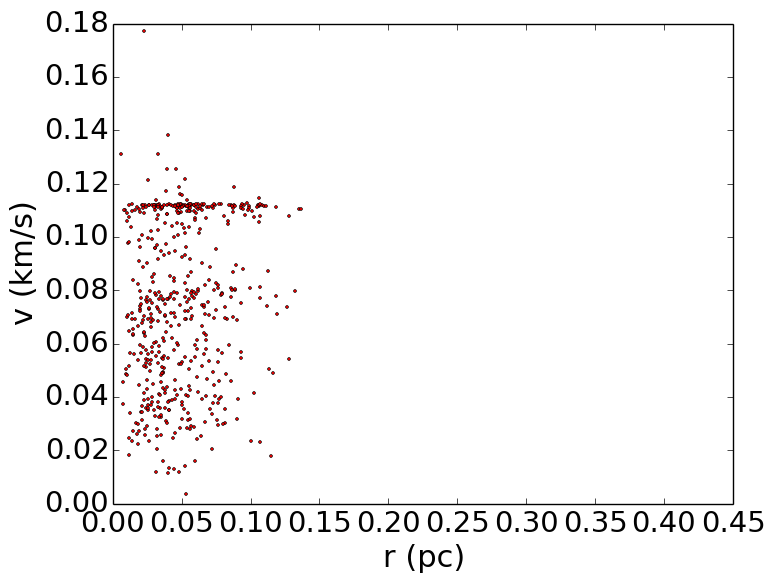} &
\includegraphics[width=0.32\textwidth,height=!]{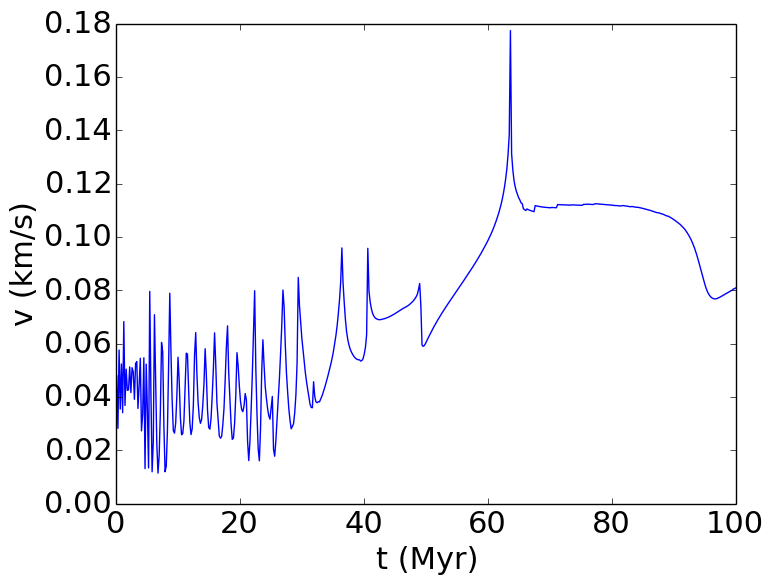} &
\includegraphics[width=0.32\textwidth,height=!]{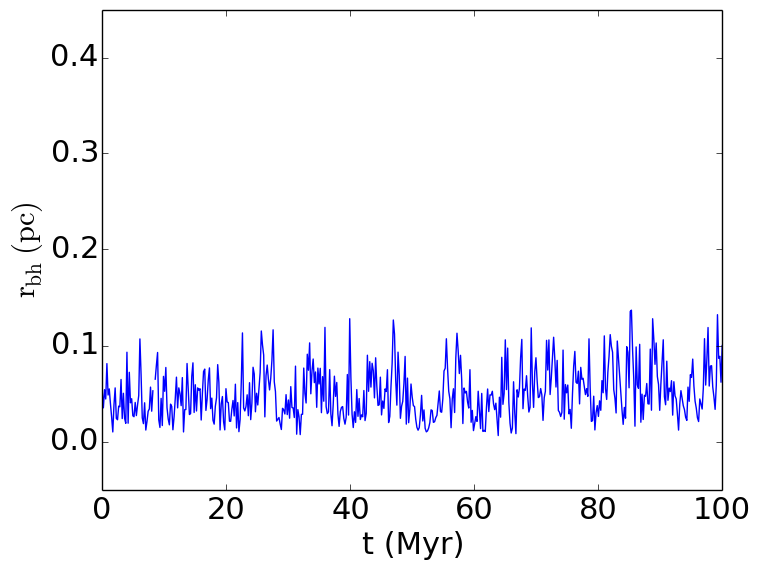} \\
\end{tabular}
\caption{Kinematic properties of the \bh{} over time, for models C50M100 ({\em top}) and C50M200 ({\em bottom}). {\em Left}: velocity versus distance from the star cluster centre; {\em middle}: velocity versus time; {\em right}: distance from the star cluster centre versus time.  \label{figure:bhvandr}}
\end{figure*}

The kinematics of an \bh{} in a star cluster strongly affects the stellar and planetary populations. Shortly after we initialise the \bh{} at rest at the cluster centre, the \bh{} experiences recoils from gravitational interactions with other stars in the core of the star cluster. Figure~\ref{figure:bhvandr} shows the kinematic properties of the \bh{} in models C50M100 and C50M200 over time. The distance and speed of the \bh{} are measured with respect the density centre of the star cluster.

The \bh{}s in models C50M100 and C50M200 spend most of their time at respective distances of 0.1~pc and 0.05~pc from their cluster centres. Larger excursions are seen in model C50M100, where the \bh{} is expelled two to a distance of $\sim 0.4$~pc from the cluster centre. These larger distances occur as a consequence of gravitational interactions with other massive objects in the star cluster. The \bh{} in model C50M200 remains closer to the cluster core that that in model C50M100, due to the differences in inertia.

The typical speed of the \bh{} during the simulation is 0.059~\kms{} in model C50M100 and 0.075~\kms{} in model C50M200. At early times ($t \la 10$~Myr) the star cluster is still compact and stellar encounters occur frequently, limiting the excursions in velocity of the \bh{}. As these encounters become less frequent due to cluster expansion after 10~Myr, the \bh{} is mostly subjected to the gravitational interaction with stars in the innermost regions of the star cluster, causing larger variations in its velocity. In model C50M200, there is a notable interaction with a binary companion, another $68~\msun$ stellar-mass black hole at $t \ga 60$~Myr.

The tidal influence of an \bh{} on planetary systems in the star cluster depends on the mass of the \bh{}, but also strongly on their mutual distance. Whereas model C50M100 contains an \bh{} that is only half the mass of that in C50M200, the motion of this \bh{} ensures that also approaches many of the stars in the region beyond the core. The \bh{} in model C50M100 experiences substantial scattering during the first few million years. Therefore, it only partially dominates the core structure, and is more subjected to stellar encounters.  The latter can explain that the  fraction of ejected planets in model C50M200 is slightly smaller than that in C50M100.  

%%%%%%%%%%%%%%%%%%%%%%%%%%%%%%%%%%%%%
%%%%%%%%%%%%%%%%%%%%%%%%%%%%%%%%%%%%%
%%%%%%%%%%%%%%%%%%%%%%%%%%%%%%%%%%%%%

\subsection{Kinematics of planetary systems and free-floating planets}

\subsubsection{Evolution of planetary systems} 

\begin{figure*}
\centering
\begin{tabular}{ccc}
  \includegraphics[width=0.32\textwidth,height=!]{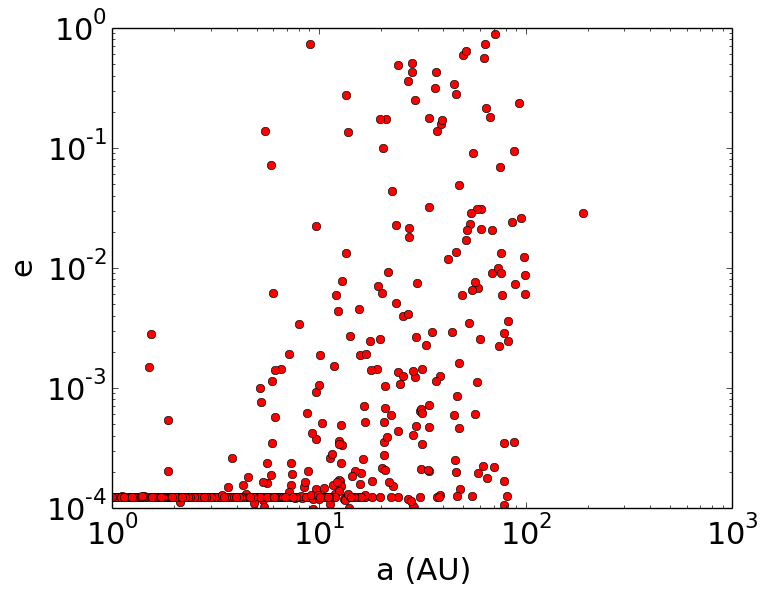} &
  \includegraphics[width=0.32\textwidth,height=!]{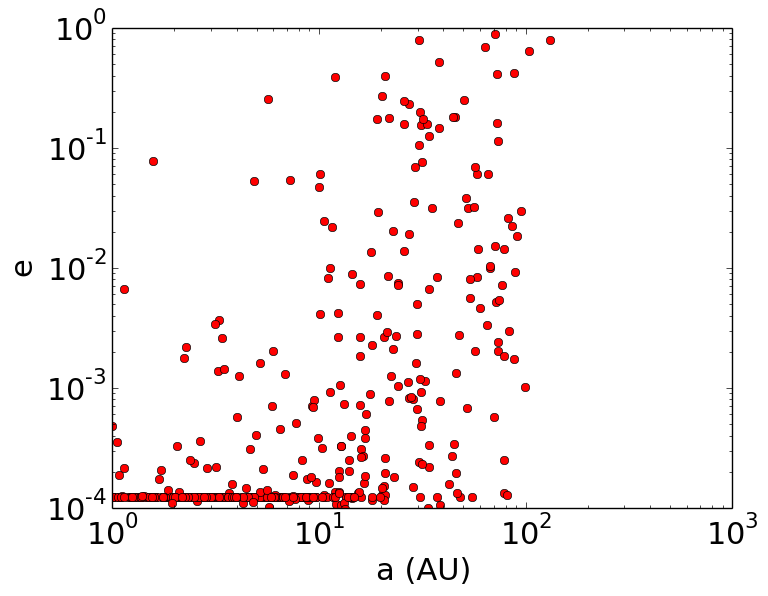}  &
  \includegraphics[width=0.32\textwidth,height=!]{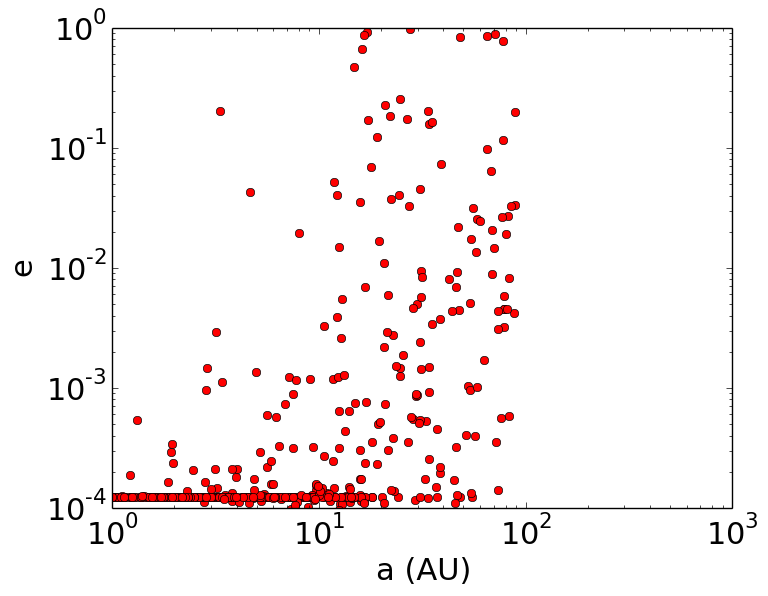}   \\
\end{tabular}
\caption{Correlations between semi-major axis and eccentricity, of all surviving planetary systems in the star cluster at $t=100$~Myr, for models C50M000 ({\em left}), C50M100, ({\em middle}) and C50M200 ({\em right}). }
\label{figure:aeiorb}
\end{figure*}

\begin{figure*}
\centering
\begin{tabular}{ccc}
  \includegraphics[width=0.32\textwidth,height=!]{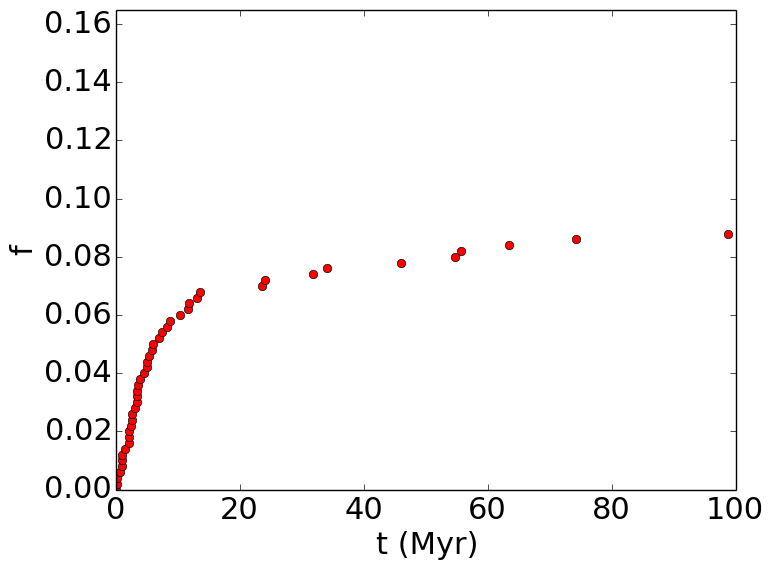} &
  \includegraphics[width=0.32\textwidth,height=!]{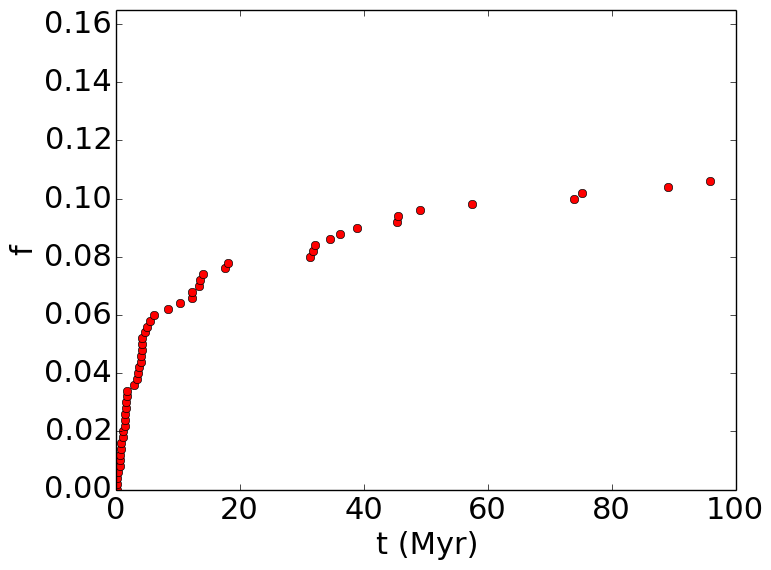}  &
  \includegraphics[width=0.32\textwidth,height=!]{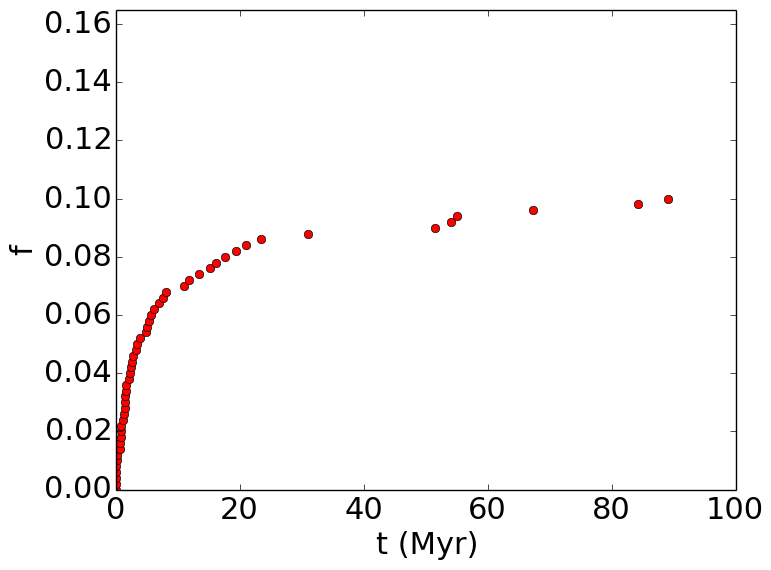}  
\end{tabular}
\caption{The cumulative fraction \bfp{}/\bbp{}$_0$ during the simulation for models C50M000 ({\em left}), C50M100, ({\em middle}) and C50M200 ({\em right}).}
\label{figure:bfpcf}
\end{figure*}

\begin{figure*}
\begin{tabular}{ccc}
  \includegraphics[width=0.33\textwidth,height=!]{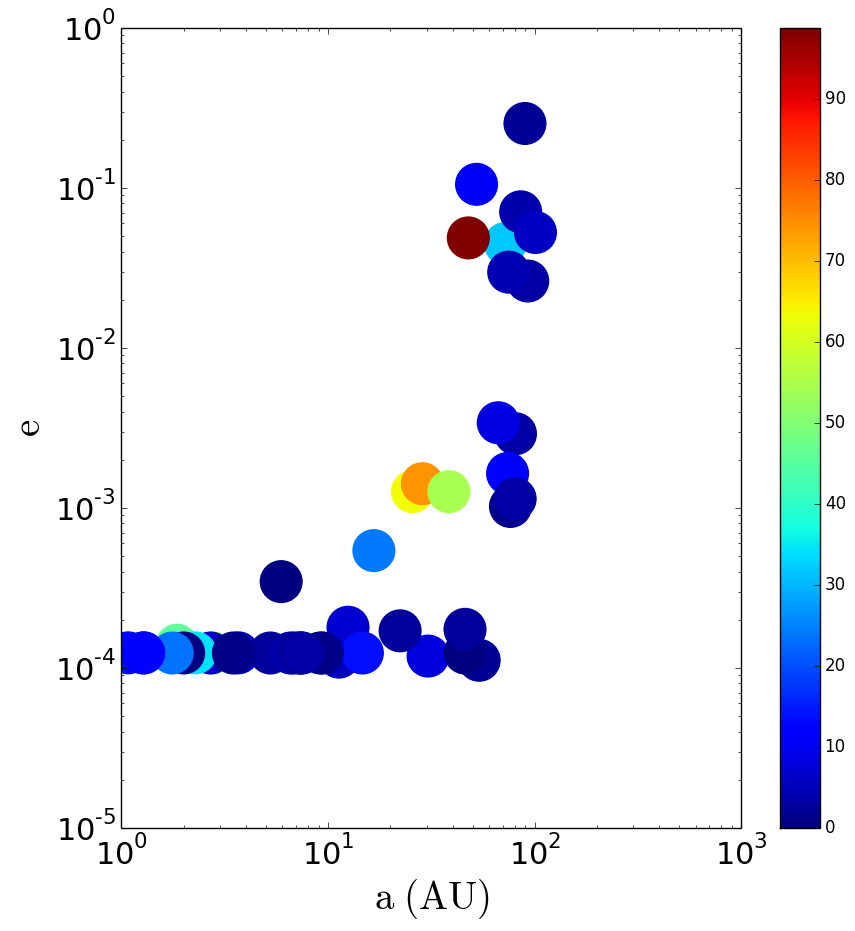} 
  \includegraphics[width=0.33\textwidth,height=!]{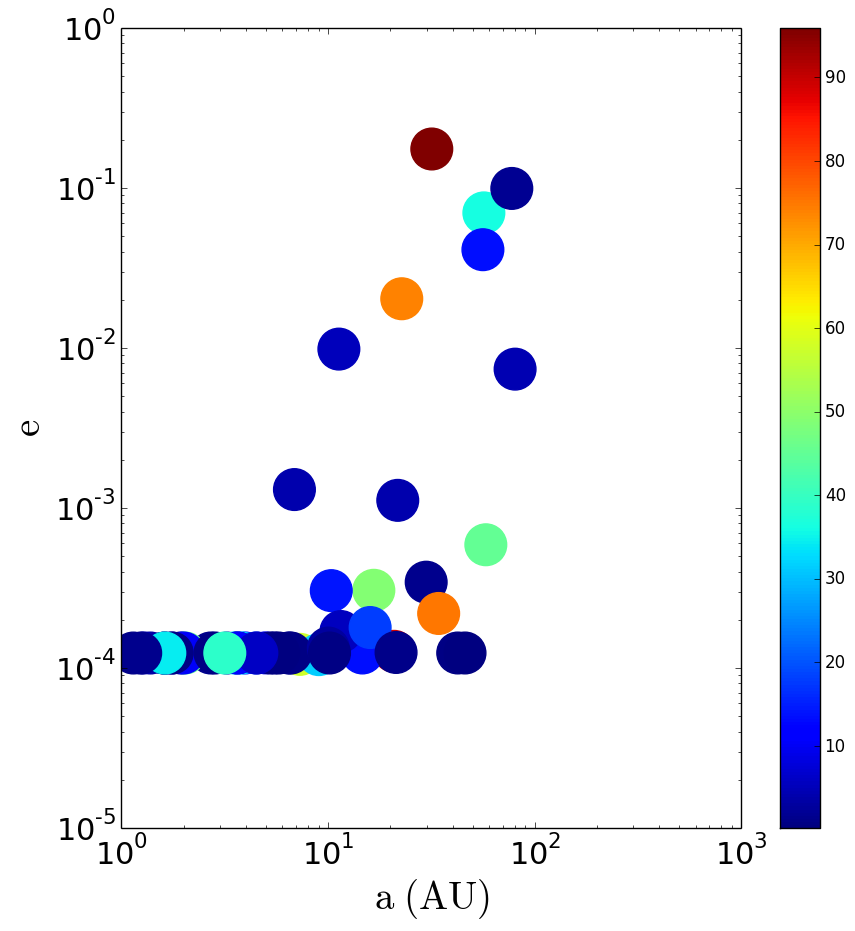} 
  \includegraphics[width=0.33\textwidth,height=!]{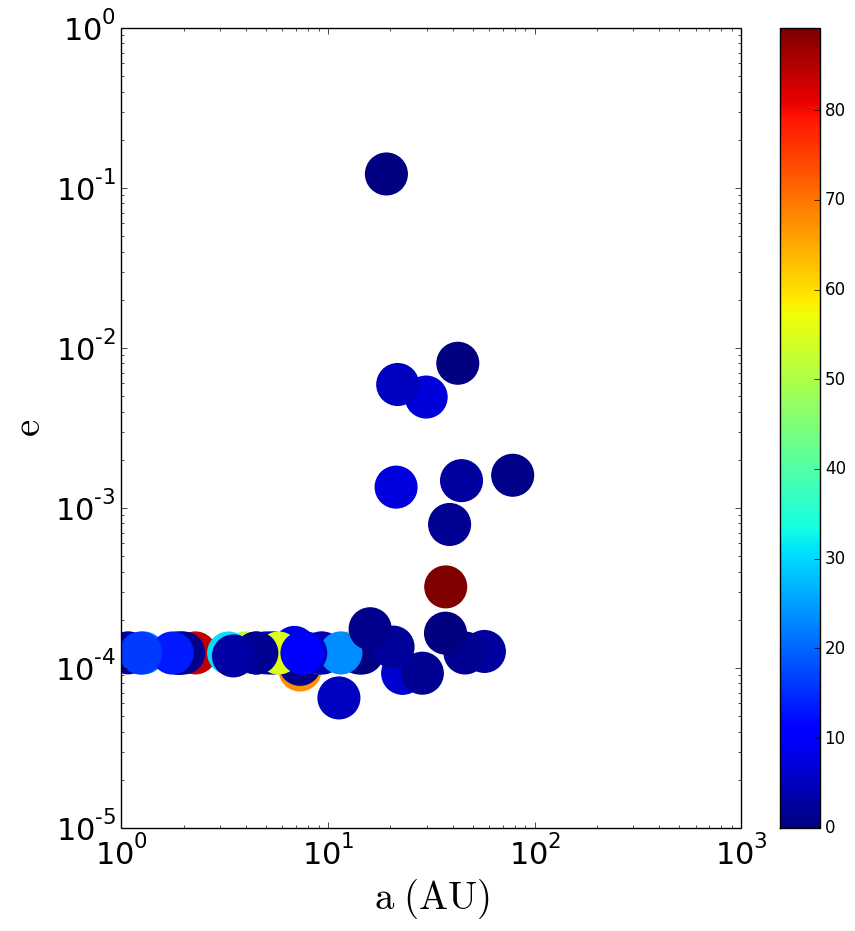}  
\end{tabular}
\caption{Semi-major axes and eccentricities of all disrupted planetary systems (the \bbp{}s), immediately before the the disruption event. The colour of each dot represents the disruption time in units of Myr. {\em Left:}  model C50M000; {\em middle:} model C50M100; {\em right:} model C50M200.}
\label{figure:bfporb}
\end{figure*}

\begin{table}
\caption{Fraction of the planetary systems in the star cluster that has remained intact (column~2), and the ratio of total number of free-floating planets, relative to the initial number of free-floating planets (column~3), at $t=100$~Myr. }
\label{tab:table}
\begin{tabular}{lcc}
\hline
Model ID  & \bbp/\bbp$_0$ & (\ffp+\bfp)/\ffp$_0$  \\
 & \% & \% \\
\hline
avC50M000 & $88.24 \pm 0.04$ & $105.95 \pm 0.03$ \\
avC50M100 & $ 84.96\pm 0.04$ & $106.38 \pm 0.03$   \\
avC50M200 & $85.14 \pm 0.03$ & $103.48 \pm 0.03$  \\
\bottomrule
\end{tabular}
\label{table:summ}
\end{table}

Small perturbations of outer planets in multi-planet systems can induce scattering events that result in the ejection of one or more planets from the system. As we only model single-planet systems in the present study, disruption of planetary systems through planet-planet scattering is not possible. Disruptions are thus always a direct consequence of (i) close encounters with neighbouring stars, and (ii) tidal interactions with the \bh{}. The majority of the planetary systems remains intact during the entire simulation, but their orbits can be significantly perturbed. Figure~\ref{figure:aeiorb} shows the correlations between the semi-major axis and eccentricity of planetary systems that have remains intact at $t=100$~Myr. As expected, the planets in long-period orbits typically experience the strongest perturbations. A comparison between the panels reveals that the direct effect of a central \bh{} in the star cluster on the orbital elements of the \bbp{}s at $t=100$~Myr is not easily distinguishable in the population of surviving planetary systems. However, the \bh{} does leave an imprint in the survival fraction, and the population of disrupted planetary system, as discussed below.

Figure~\ref{figure:bfpcf} shows the cumulative fraction of the times at which \bfp{}s appear in the simulations (i.e., when planetary systems are disrupted). Most disruptions occur within one relaxation time, when the star cluster is compact and when the \bh{} strongly influences the dynamics of both the stellar and the planetary populations.
We estimate that a $1\msun$ star  in  model C50M000 undergoes an encounter within a distance of $1000$~AU every $t_{\rm enc} \approx 5.3$~Myr. The corresponding encounter frequencies are 5.1~Myr and 4.9~Myr for models C50M100 and C50M200, respectively \citep[see also][]{malmberg2007}. Of the total number of planetary systems disrupted during the entire simulation (i.e., $100$~Myr), half are disrupted within the first $\sim 6$~Myr, $\sim 6$~Myr, and $\sim 4$~Myr for models C50M000, C50M100 and C50M200,  respectively.

Table~\ref{table:summ} shows that in the models that contain an \bh{}, the number of planetary systems in the star cluster is reduced by approximately 16\%, due to escape of planetary systems from the star cluster ($\approx 4\%$), and due to disruptions of planetary systems ($\approx 12\%$). In model without a central \bh{}, on the other hand, the number of planetary systems is reduced by approximately only 11\%.
In models containing an \bh{}, mass segregation quenching prevents the star cluster core from losing the majority of its members during the early phases of the star cluster dynamical evolution (see Figure~\ref{figure:chnandm}). 
After the first $10$~Myr, the star cluster expands and starts to slowly fill its Roche Lobe. At $t \approx \relaxationtime$, the disruption rate of planetary systems drops significantly, and the three models have a similar, more or less constant ejection rate.

Figure~\ref{figure:bfporb} shows the orbital elements of the planetary systems at the time-step prior to disruption of the system, with colours indicating the time of ejection. There is a non-negligible percentage of $1-10$~AU planets ejected from their host stars, while the $10-100$~AU region is of more difficult interpretation, as the star cluster density can strip those planets easier. The majority of these, which semi-major axis is below 10~AU, escape before 10~Myr, while most of the \bfp{} escape before 30~Myr. 
The presence of a central \bh{} in the star cluster enhances the fraction of \bfp{}s originating from systems with $a=1-10$~AU by $13-16\%$, as compared to the model without an \bh{}.
Therefore, a central \bh{} enhances the disruption of short-period planets. The total number of \bfp{}, coming from the semi-major axis interval $1-10$~AU, is enhanced by $\sim 2\%$ from model C50M000 to C50M100. These results suggest that planetary systems with small semi-major axes are disrupted more easily in star clusters containing an \bh{} as compared to in star clusters without an \bh.

The third column of Table~\ref{table:summ} shows the ratio of the total number of free-floating planets (\bfp{}+\ffp{}) in the cluster, as compared to its initial value (\ffp$_0$), at $t=100$~Myr. The number of \bfp{}s is higher for models containing an \bh{}, due to the disruption of planetary systems. The ejection rate of \ffp{}s, on the other hand, is also higher when a central \bh{} is present. Hence, the highest number of free-floating planets (\bfp{}+\ffp{}) can be found in model C50M100.

%%%%%%%%%%%%%%%%%%%%%%%%%%%%%%%%%%%%%%%%%%%%%%%%%%%%%%%
%%%%%%%%%%%%%%%%%%%%%%%%%%%%%%%%%%%%%%%%%%%%%%%%%%%%%%%
%%%%%%%%%%%%%%%%%%%%%%%%%%%%%%%%%%%%%%%%%%%%%%%%%%%%%%%

\subsubsection{Density profiles and \bh{} effects on planetary systems \label{sec:density}}

\begin{table*}
\caption{The fraction of the planets in the star that have escaped from their host stars that is located in the star cluster (\bfp{}) or outside the star cluster (e\bfp{}), and fraction of the initially free-floating planets that has escaped from the star cluster, for star clusters with different initial masses, at $t=100$~Myr. The density model (d$n$) is indicated at the top of each column, with "d0", "d1", and "d2" for $\nstars=5000$, 10\,000, and 15\,000, respectively.
\label{tab:table}}
\begin{tabular}{llcccc}
\hline
ID  & Quantity  & d0 ($\nstars=5\,000$) & d1 ($\nstars=10\,000$)  & d2 ($\nstars=15\,000$)   \\
\hline
C50M000d$n$ & \bfp/\bbp$_0$ & $4.8 \pm 0.9 \ \%$  & $9.9 \pm 1.3 \ \%$  & $12.8 \pm 1.5 \ \%$ \\
 &  e\bfp/\bbp$_0$ &  $1.0 \pm 0.4 \ \%$  &  $1.3 \pm 0.5 \ \%$ &  $1.4 \pm 0.5 \ \%$ \\
 & e\ffp/\ffp$_0$ & $3.2 \pm 0.8 \ \%$ & $2.2 \pm 0.6 \ \%$  & $2.2 \pm 0.7 \ \%$ \\
 & $r_c$ & 0.34~pc & 0.39~pc & 0.29~pc \\
 & $\halfmass$  & 0.80~pc & 0.78~pc & 0.80~pc \\
 \hline
C50M100d$n$ & \bfp/\bbp$_0$ & $6.0 \pm 1.1 \ \%$  & $11.5 \pm 1.4 \ \%$  & $16.0 \pm 1.6 \ \%$  \\
 & e\bfp/\bbp$_0$ &  $2.0 \pm 0.6 \ \%$ &  $1.7 \pm 0.6 \ \%$ & $2.2 \pm 0.7 \ \%$ \\
 & e\ffp/\ffp$_0$ & $5.0 \pm 1.0 \ \%$  & $3.6 \pm 0.8 \ \%$  & $2.2 \pm 0.7 \ \%$ \\
 & $r_c$ & 0.19~pc & 0.24~pc & 0.15~pc \\
 & $\halfmass$  & 0.83~pc & 0.79~pc & 0.81~pc \\
  \hline
C50M200d$n$ & \bfp/\bbp$_0$ & $8.2 \pm 1.2 \ \%$  & $11.2 \pm 1.4 \ \%$  & $16.2 \pm 1.6 \ \%$  \\
 &e\bfp/\bbp$_0$ &  $1.4 \pm 0.5 \ \%$ &  $2.2 \pm 0.6 \ \%$ &  $2.6 \pm 0.7 \ \%$ \\
 & e\ffp/\ffp$_0$ & $5.6 \pm 1.0 \ \%$  & $5.5 \pm 1.0 \ \%$  & $3.0 \pm 0.8 \ \%$  \\
 & $r_c$ & 0.11~pc & 0.15~pc & 0.09~pc \\
 & $\halfmass$  & 0.86~pc & 0.80~pc & 0.82~pc \\
\bottomrule
\end{tabular}
\label{table:densitychange}
\end{table*}

\begin{figure}
  \includegraphics[width=0.52\textwidth,height=!]{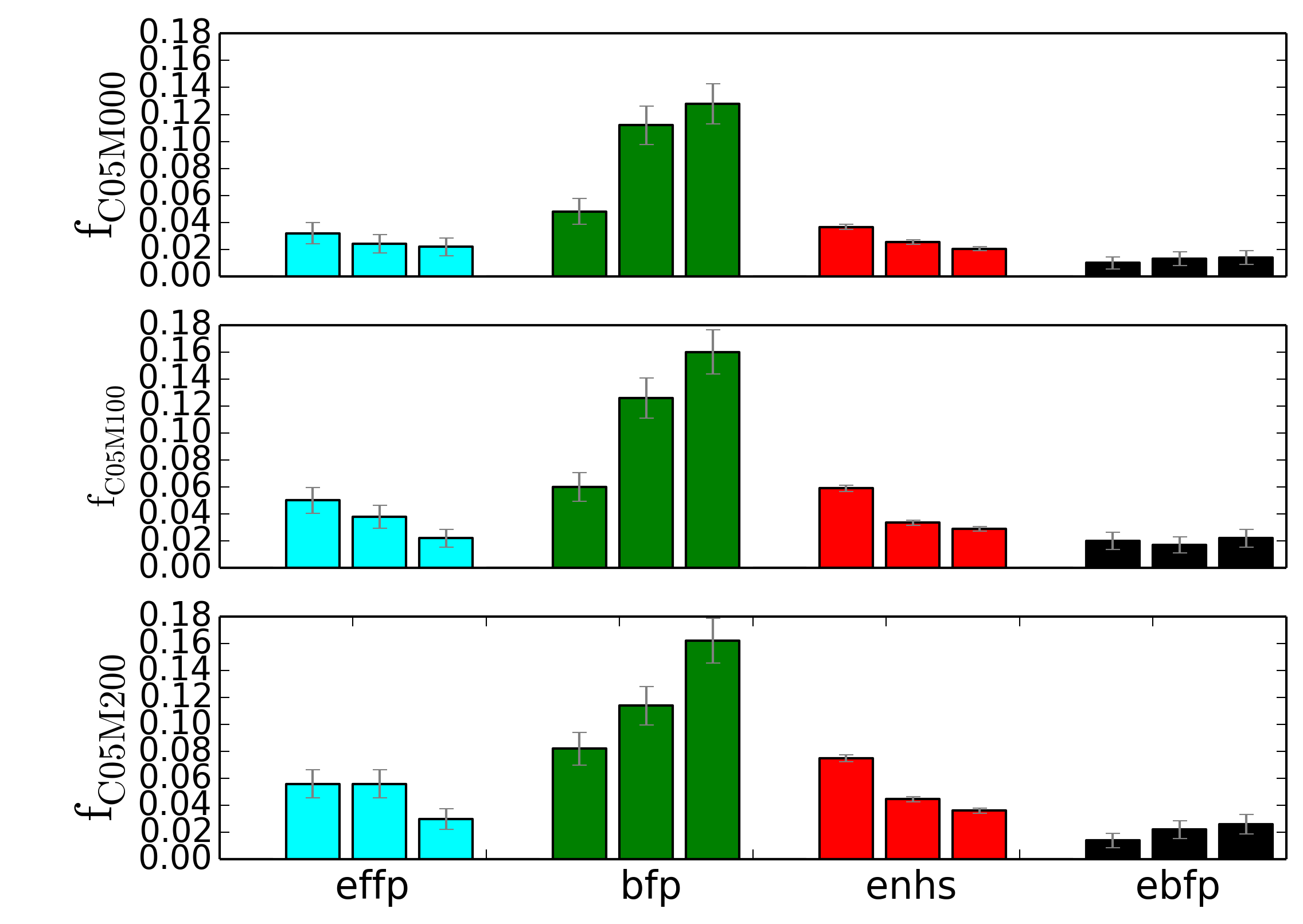}  
\caption{Fractions e\ffp/\ffp$_0$, \bfp/\bbp$_0$, e\nhs{}/\nhs{}$_0$  and e\bfp{}/\bbp$_0$ escaped from the star cluster fraction, at $t=100$~Myr, for models C50M000 ({\em top}), C50M100 ({\em middle}) and C50M200 ({\em right}). }
\label{figure:densdist}
\end{figure}

\begin{figure*}
\begin{tabular}{cc}
 \includegraphics[width=0.4\textwidth,height=!]{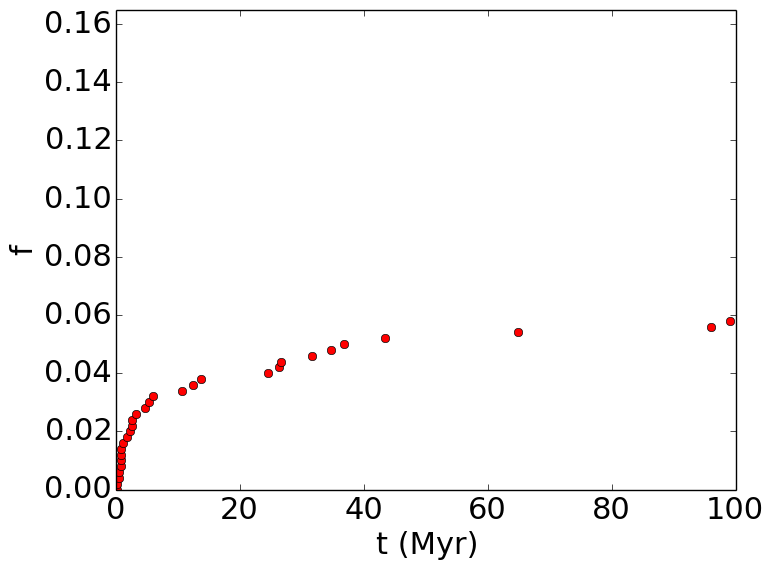} &
 \includegraphics[width=0.4\textwidth,height=!]{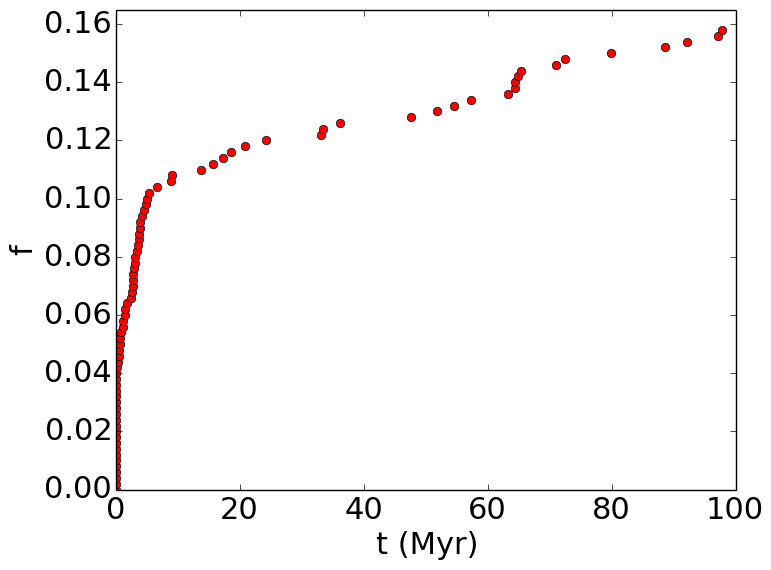} \\
 \includegraphics[width=0.4\textwidth,height=!]{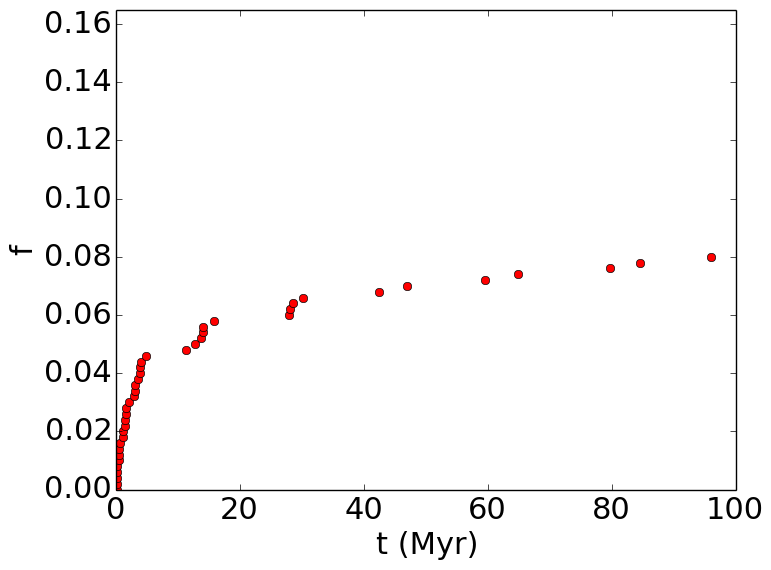} &
 \includegraphics[width=0.4\textwidth,height=!]{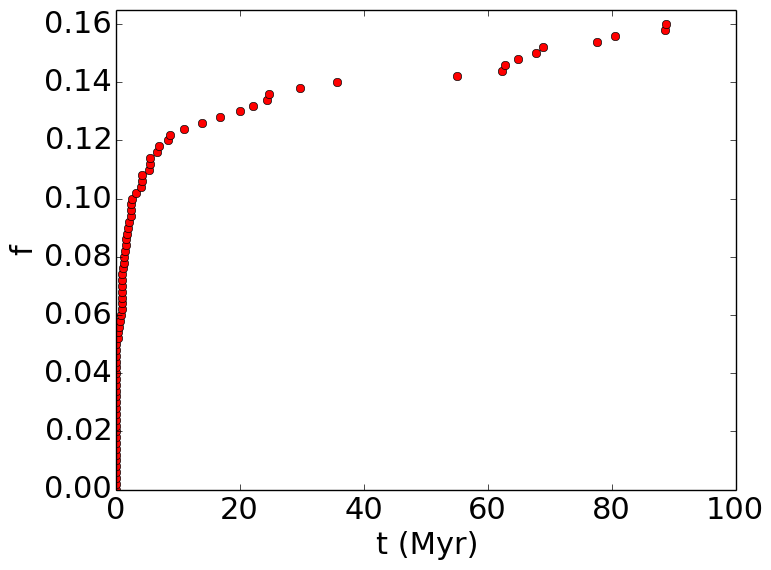} \\
\end{tabular}
\caption{The cumulative fraction \bfp{}/\bbp{}$_0$ generated in the star cluster, for models C50M100d0 ({\em top-left}), C50M200d0 ({\em bottom-left}), C50M100d2 ({\em top-right}) and C50M200d2 ({\em bottom-right}). }
\label{figure:densefigure}
\end{figure*}

The stellar density is among the most important parameters that determines the stability of planetary systems. To investigate the effect of stellar density, we carry out simulations of star clusters with different initial stellar densities. We carry out simulations with $\nstars=5\,000$, 10\,000 and 15\,000 stars, while retaining the same initial half-mass radius as in our reference model. We label these models with the template C-Q0-\bh{}-mass, with a postfix "d0", "d1" or "d2", corresponding to the models with $\nstars=5\,000$, 10\,000 and 15\,000 respectively.

As can be seen in Table~\ref{table:densitychange} and Figure~\ref{figure:densdist}, the number of \bfp{}s increases with increasing stellar cluster density. Mass segregation is quenched in all models containing an \bh{}. However, a larger stellar density results in a higher rate of close encounters between stars. 
The total mass star clusters C50M000d0, C50M100d0, and C50M200d0 is approximately half of the mass of the corresponding reference models. 
Due to their shallower gravitational potential well, these clusters have, on average, a higher production rate of e\ffp{}s. 
 The number of ejected stars is numerically larger in denser star clusters (since $\nstars$ is larger), although the fraction of stars that is ejected from the star clusters is lower. For the same reason,  models C50M000d2, C50M100d2, and C50M200d2, are similarly influenced by the \bh{} in its core and the intermediate regions within the $\halfmass$, which becomes more populated.

Figure~\ref{figure:densefigure} provides us important insights into the disruption mechanisms of planetary system in a more and less dense star cluster. When comparing Figures~\ref{figure:bfpcf} and~\ref{figure:densefigure}, we can see that the fraction of \bfp{} depends strongly on both the presence of an \bh{}, as well as the star cluster density. 
Half of the planetary systems are disrupted before 7~Myr for model C50M100d0 and 6~Myr for model C50M200d0. The same fraction of planetary systems is disrupted in less than 1~Myr in the denser movels C50M100d2 and C50M200d2. The star cluster density and \bh{} tidal effect are intrinsically connected. The \bh{}'s tidal field enhances the stellar velocities in its neighbourhood, and in a denser star cluster, encounters occur more frequently, and there are more ejections than in star clusters with a lower stellar density.

%%%%%%%%%%%%%%%%%%%%%%%%%%%%%%%%%%%%%%%%%%%%%%%%%%
%%%%%%%%%%%%%%%%%%%%%%%%%%%%%%%%%%%%%%%%%%%%%%%%%%
%%%%%%%%%%%%%%%%%%%%%%%%%%%%%%%%%%%%%%%%%%%%%%%%%%

\subsubsection{Dependence on initial semi-major axis distribution \label{auchange}}

\begin{table*}
\caption{Fractions of \bfp{}s and e\bfp{}s for models with different initial semi-major axis intervals, at $t=100$~Myr. The models are indicated in the first column, with a$n$ indicating the initial semi-major axis interval ("a0", "a1", and "a2" for $1-10$ AU, $1-100$~AU and $10-100$`AU, respectively) in columns $2-7$. Columns~8 and~9 lists the core radii and half-mass radii, respectively.
\label{tab:table}}
\begin{tabular}{lcccccccc}
\hline
Model ID & \multicolumn{2}{c}{a0 ($1-10$~AU)}  &  \multicolumn{2}{c}{a1 ($1-100$~AU)} & \multicolumn{2}{c}{a2 ($10-100$~AU)}  & $r_c$ & $\halfmass$ \\
   & \bfp/\bbp$_0$ & e\bfp/\bbp$_0$ & \bfp/\bbp$_0$ & e\bfp/\bbp$_0$ & \bfp/\bbp$_0$ & e\bfp/\bbp$_0$ & pc & pc   \\
\hline
C50M000a$n$ & $0.6 \pm 0.3 \ \%$ & $0.2 \pm 0.2 \ \%$  & $9.9 \pm 1.3 \ \%$ & $1.3 \pm 0.5 \ \%$ & $21.4 \pm 1.8 \ \%$ & $2.0 \pm 0.6 \ \%$ & 0.39 & 0.78 \\
C50M100a$n$ & $1.2 \pm 0.5 \ \%$ & $0.4 \pm 0.3 \ \%$ & $11.6 \pm 1.4 \ \%$ & $1.7 \pm 0.6 \ \%$ & $21.0 \pm 1.8 \ \%$ & $2.8 \pm 0.7 \ \%$ & 0.24 & 0.79 \\
C50M200a$n$ & $1.4 \pm 0.5 \ \%$ & $0.4 \pm 0.3 \ \%$ & $11.2 \pm 1.4 \ \%$ & $2.2 \pm 0.6 \ \%$ & $23.2 \pm 1.9 \ \%$ & $4.2 \pm 0.9 \ \%$  & 0.15 & 0.80 \\
\bottomrule
\end{tabular}
\label{table:smachange}
\end{table*}

\begin{figure}
\centering
\begin{tabular}{c}
 \includegraphics[width=0.4\textwidth,height=!]{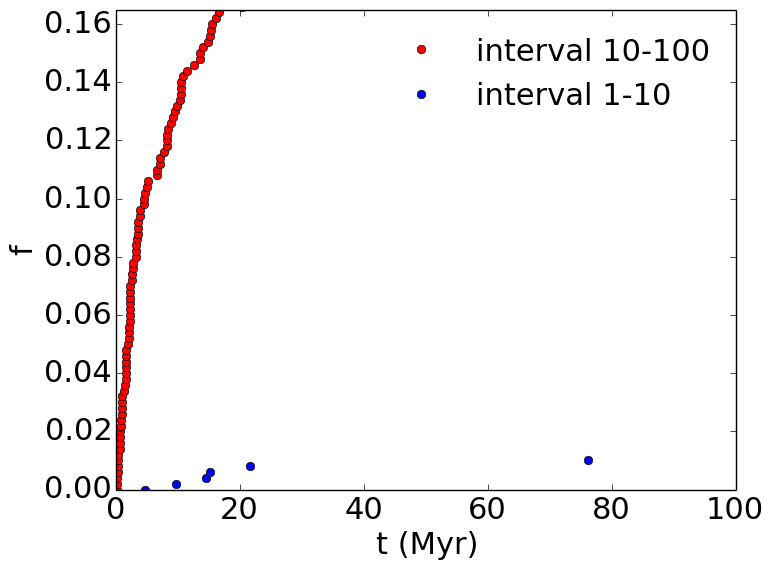} \\
 \includegraphics[width=0.4\textwidth,height=!]{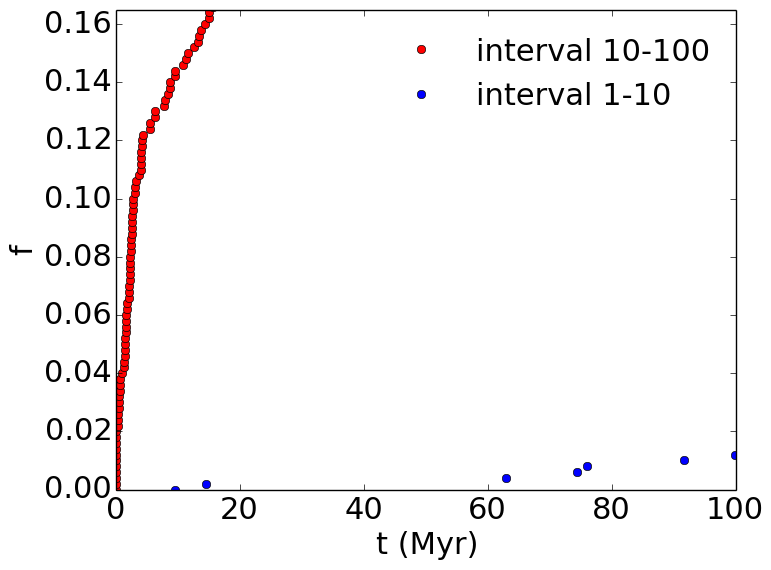} \\
\end{tabular}
\caption{The cumulative fraction \bfp{}/\bbp{}$_0$ for models C50M100a0 and C50M100a2 ({\em top}) and for models C50M200a0 and C50M200a2 ({\em bottom}).}
\label{figure:audiff}
\end{figure}

\begin{figure}
  \includegraphics[width=0.5\textwidth,height=!]{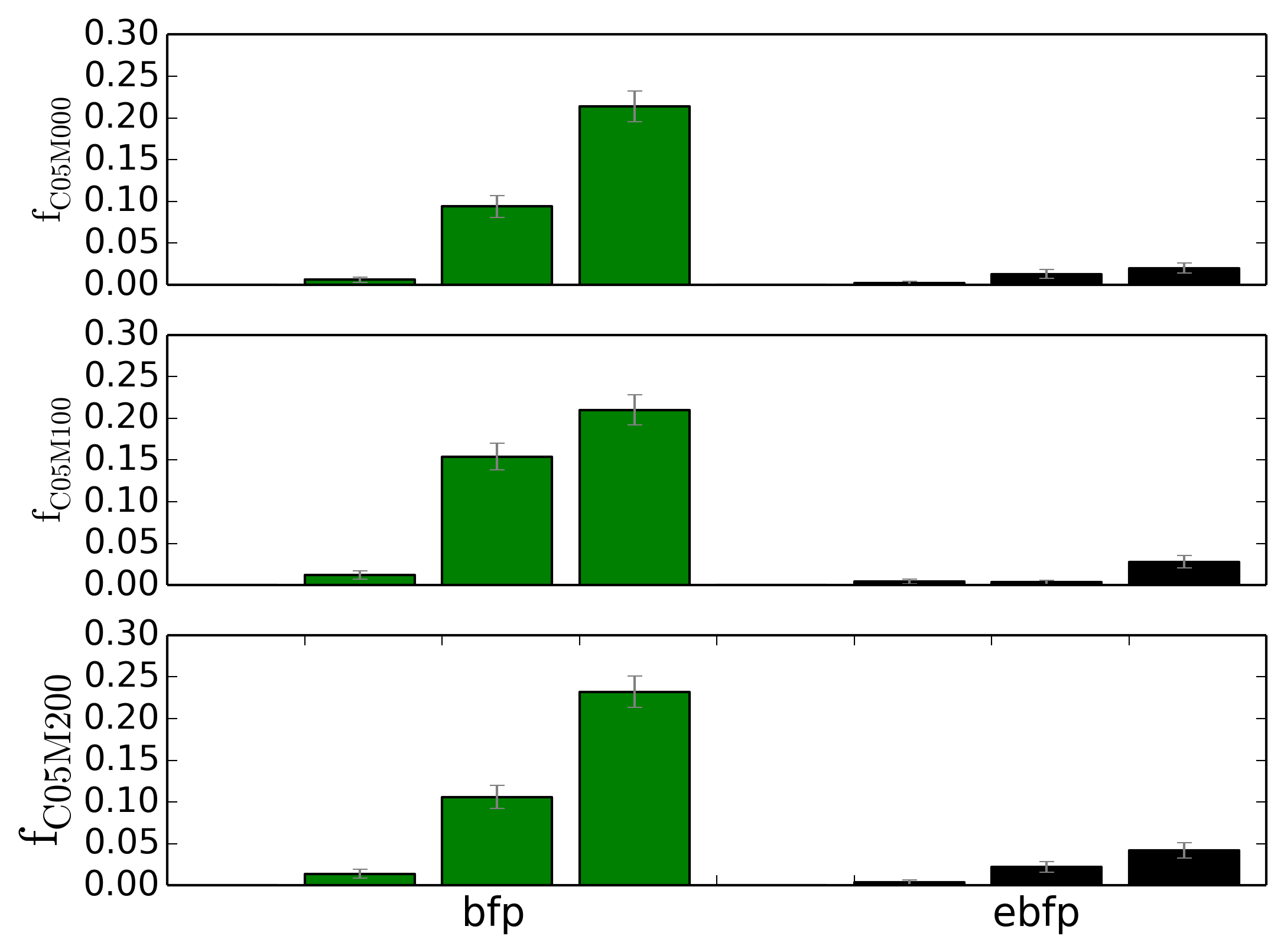}  
\caption{Fractions \bfp{}/\bbp{}$_0$ (green bars) and e\bfp{}/\bbp{}$_0$ (black bars) at $t=100$~Myr, for different initial semi-major axis distributions. From left to right, the bars in each group represent initial semi-major axis distributions in the range $1-10$~AU,  $1-100$~AU, and $10-100$~AU, respectively, for models C50M000 ({\em top}), C50M100 ({\em middle}) and C50M200 ({\em bottom}). }
\label{figure:audist}
\end{figure}

We study how the star-planet systems with different semi-major axes evolve, in order to understand the implications of the \bh{} contribution for the evolution and distruption rates of short-period planetary systems. We study star-planet systems with semi-major axes distributions drawn from uniform log-normal distributions in the range $1-10$~AU (models "a0"), $1-100$~AU (models "a1" as in the reference models) and $10-100$~AU (models "a2"); see Table~\ref{table:smachange}. In star clusters similar to those in our simulations, the  critical value of $a=10$~AU separates two different probability ejection regimes \citep[e.g.,][]{fujii2019, flammini2019}.

The blue dots in Figure~\ref{figure:audiff} represent the $1-10$~AU semi-major axis interval, and are consistent with the results of \cite{flammini2019} and \cite{fujii2019}. The fraction of disrupted planetary systems is again larger than in the model without an \bh{}, by 100\% and 133\% in models C50M100a0 and C50M200a0, respectively. 
The red dots in Figure~\ref{figure:audiff} represent $10-100$~AU semi-major axis interval, and exhibit a similar behaviour as in Figure~\ref{figure:densefigure}. Planets with larger semi-major axes have a larger probability of being ejected both due to stellar encounters, combined with the tidal influence of the \bh{}. It is difficult to disentangle the relative contributions of these two processes.
In Table~\ref{table:smachange} and Figure~\ref{figure:audist}, we obtain similar results as \cite{fujii2019,flammini2019}. When comparing the different models, the number of disrupted planetary systems (\bfp{}s) is large for the $10-100$~AU interval, while it remains small for the $1-10$~AU. Planetary systems with $a \la 10$~AU (Saturn's orbit) are relatively difficult to disrupt, unless these are multi-planet systems, in which planet-planet scattering can occur. 
The necessary conditions related to velocity of both host star and encountering star are more likely to be reached inside the core, rather than outside. On the $10-100$~AU interval we have a different situation. Planets are far more easy to be ejected, also trough average close encounter.

%%%%%%%%%%%%%%%%%%%%%%%%%%%%%%%%%%%%%
%%%%%%%%%%%%%%%%%%%%%%%%%%%%%%%%%%%%%
%%%%%%%%%%%%%%%%%%%%%%%%%%%%%%%%%%%%%

\section{Discussion and conclusions} \label{section:conclusions}

We have numerically investigated of the influence of a central \bh{} on the dynamical evolution of planetary systems and free-floating planets in star clusters. We provide analytic estimates for the tidal force exerted on planetary systems by the \bh{}, as compared to that of the nearest neighbour stars. We characterise survival rates of planetary systems, and escape rates of stars, planetary systems, and free-floating planets, as a function of time, \bh{} mass, the initial stellar density, and the initial planetary semi-major axis distribution. Our findings can be summarised as follows.

\begin{enumerate}

\item Planetary systems experience a tidal force of neighbouring stars and of the central \bh{}. Depending on the density profile of the star cluster, the \bh{}'s tidal force dominates over that of nearest neighbour stars in large regions of the star cluster. In the Plummer model, a central \bh{} with a mass larger than 3.4\% of the total stellar mass dominates the tidal force throughout the entire star cluster. A central \bh{} exerts a continuous tidal force on all planetary systems in the star cluster, while close encounters with nearby neighbours can have substantial short-term effects. 
Finally, it is rather unlikely that the \bh{} will directly strip the host star of its planet. The \bh{} is in the core, therefore there is a larger probability of an encounter that destroy the planetary system before it reaches the \bh.

\item Gravitational interactions with neighbouring stars affect planetary systems in star clusters, and the \bh{} enhances these effects. Moreover, we find that the \bh{} has the most prominent effect during in the early and more compact phases of the star cluster life. The escape rate of both stars and planets in the star cluster is higher when a IMBH is present in the star cluster.

\item The quenched mass segregation mechanism modifies the evolution of the star cluster, in the presence of a central \bh{}. The stars are bounced back from the star cluster core, resulting in a relatively poor core population, from which the more massive stars are also regularly ejected, and more stars in the intermediate region within the $\halfmass$.

\item The kinematics of free-floating planets (\bfp{}s and \ffp{}s) is mostly affected by the gradual change in the star cluster's global gravitational potential. The effect of a central \bh{} on stellar ejection rates and reshaping of the cluster's density profile, combined with stellar evolution, determines the ejection rates of the free-floating planets.

\item The amplitude of the motion of the \bh{} in the star cluster core due to interactions with stars in the cluster core is smaller  for higher-mass \bh{}s. As the location of the \bh{} in the star cluster determines the dynamics of the neighbouring population, and also the tidal effect on planetary systems, this behaviour may justify the slightly larger number of disrupted planetary systems in the models with a $100\msun$ \bh{}, as compared to the $200\msun$ \bh.

\item The rate of planetary system disruptions (and consequently, the fraction of \bfp{} in the cluster) is higher when an \bh{} is present. The \bh{} enhances the velocities of the stars passing near it and enhance the close encounter ratio in the star cluster. The production rates of \bfp{}s and e\ffp{}s are numerically larger in models containing a central \bh{}, as compared to star clusters without an \bh{}. The trend is confirmed for different star cluster densities and, for \bfp, for models with different initial  semi-major axis intervals as well.

\item All particles in the star cluster are directly or indirectly affected by the \bh. This affect is most important during the first $\sim 10$~Myr, until the star cluster fills its Roche lobe. The planets ejected from their host star are affected on all the tested semi-major axis. Even for planets in short-period orbits ($a<10$~AU), there is a, non-negligible, additional contribution of ejected planets when an \bh{} is present in the cluster.

\item Within a timescale of 100~Myr, star cluster with a higher initial stellar density have a higher disruption rate of planetary systems, due to the increased frequency of close encounters, and a reduced production rate of ejected free-floating planets, due to the deeper gravitational potential. 

\end{enumerate}

Our study highlights the importance of an \bh{} when considering planetary systems in star clusters. Although the presence of \bh{}s in the centres of star clusters has not yet been unambiguously determined, continued observational efforts are likely to reveal their presence in the coming decade. Aside from observational work, further computational efforts are also necessary to characterise the influence of \bh{}s with a wide range of masses on multi-planet systems in globular clusters.

%%%%%%%%%%%%%%%%%%%%%%%%%%%%%%%%%%%%
%%%%%%%%%%%%%%%%%%%%%%%%%%%%%%%%%%%% 
%%%%%%%%%%%%%%%%%%%%%%%%%%%%%%%%%%%%

\section*{Acknowledgements}

We are grateful to the anonymous referee for providing comments and suggestions that helped to improve this paper.
F.F.D. acknowledges support from the XJTLU postgraduate research scholarship.
M.B.N.K. acknowledges support from the National Natural Science Foundation of China (grant 11573004). This research was supported by the Research Development Fund (grant RDF-16-01-16) of Xi'an Jiaotong-Liverpool University (XJTLU). 
We acknowledge the support of the DFG priority program SPP 1992 "Exploring the Diversity of Extrasolar Planets (Sp\,345/20-1)". 
We thank Zhongmu Li of Dali University, Yunnan, China for the organization, kind hospitality and support during a workshop in Dali in 2019 and
Hyung Mok Lee and the Korea Astronomy and Space Science Institute (KASI) in Daejoen, Korea (Rep.), for financial support during the KCK11 meeting in Dec. 2019. This work has been partly supported by Sino-German cooperation (DFG, NSFC) under project number GZ 1284. We are grateful to Martin Gorbahn for discussions that helped to improve the paper. 

\section*{Data availability}

The data underlying this article will be shared on reasonable request to the corresponding author. \\

\bibliography{paperII.bib}
\label{lastpage}
\end{document}